\documentclass[a4paper,journal]{IEEEtran}

\usepackage{amsmath,amssymb,amsfonts,amsthm,xcolor,makeidx,graphicx}

\usepackage{bm}
\usepackage{multirow}

\usepackage{enumitem}
\usepackage{tikz}
\usetikzlibrary{arrows.meta}

\usepackage{cite}

\usepackage{array,makecell,adjustbox}
\usepackage{pifont} 
\newcommand{\cmark}{\ding{51}}

\newcolumntype{P}[1]{>{\centering\arraybackslash}p{#1}}

\usepackage[final]{microtype}
\microtypesetup{protrusion=true,expansion=true}

\usepackage[most]{tcolorbox}

\usepackage[normalem]{ulem}
\usepackage{xcolor}

\usepackage{tabularx}   
\usepackage{booktabs}   

\newtheorem{remark}{Remark}
\definecolor{lightblue}{RGB}{230, 245, 255} 
\definecolor{lightblue_title}{RGB}{245, 250, 255}  

\begin{document}

\title{Towards Network-Aware Operation of Integrated Energy Systems: A Comprehensive Review}

\author{Alessandra Parisio,~\IEEEmembership{Senior Member,~IEEE}
\thanks{Department of Electrical and Electronic Engineering,
University of Manchester, Oxford Road, Manchester M13 9PL, United Kingdom}
\thanks{This work was partially supported by the Supergen Energy Networks Impact Hub (EP/Y016114/1). \\
Preprint notice: This work has been submitted to the IEEE for possible publication. Copyright may be transferred without notice, after which this version may no longer be accessible.
\\
© 2026 Alessandra Parisio. All rights reserved. 
Figures, tables, and text may not be reused without permission.}}

\maketitle

\begin{abstract}
Integrated Energy Systems (IES) are systems of interconnected electricity, gas, heating, and cooling networks, where the carriers interact and depend on one another. Beyond these core vectors, IES may also incorporate additional infrastructures, such as hydrogen, transportation and water networks, whenever sector coupling or cross‑vector exchanges are relevant. Although modern cities already function as multi-energy systems, these networks are still planned and operated in isolation, which leads to inefficiencies and unused flexibility. As distributed energy resources (DERs) grow, local coupling among electricity, heating, and gas networks becomes stronger, so coordinated operation across carriers and infrastructures is essential. IES can improve efficiency, flexibility, and renewable integration, yet operation is challenging because of complex interdependencies, non-convex behaviors, and multi-scale dynamics of the energy networks. A key point that the literature often overlooks is the explicit role of network constraints and topology, which shape feasible operating regions, affect scalability, and determine how uncertainty and formal guarantees can be addressed. 
This review provides a first comprehensive analysis of network‑aware modeling, optimization, and control methods for IES. We identify methodological limitations related to tractability, feasibility guarantees, and scalability. Building on these insights, we outline research directions that include distributed optimization with theoretical guarantees and control approaches informed by operational data. The review offers a foundation for scalable, network-aware operational frameworks for future low-carbon energy systems.
\end{abstract}

\begin{IEEEkeywords}
Integrated Energy System, Modeling, Optimization, Model Predictive Control.
\end{IEEEkeywords}

\section{Integrated Energy Systems: A Backbone in the Energy Transition}\label{MEN}

\IEEEPARstart{T}he global energy transition, driven by decarbonization, decentralization, and digitalization, is reshaping how energy systems are designed and operated~\cite{IRENA2024}. Integrated Energy Systems (IES) have emerged as a key framework for coordinating electricity, gas, heating, and cooling networks to improve efficiency, flexibility, and sustainability, and are particularly critical at the distribution level, where local coordination improves energy balance and operational performance. 

Simulations show potential primary energy savings of up to 53.5\% and CO$_{2}$ reductions of up to 76\%~\cite{Xu2024}. However, these promising outcomes are based on simplified models that neglect the behavior and constraints of the underlying energy networks. For realistic assessments, network behavior and physical constraints, such as flow limits, pressure drops, and thermal losses, must be incorporated~\cite{7842813,Malley2020}. These network characteristics fundamentally shape feasible operating regions and determine the flexibility that can be exploited in practice.

Managing flexibility is central to integrating variable renewables, reducing operational costs, and improving system-wide resilience. Flexibility stems from storage, controllable demand, conversion technologies such as power-to-gas and power-to-heat, and aggregation of distributed energy resources (DERs)~\cite{Chicco2020}. Yet, integrated operation enables cross-vector flexibility only to the extent permitted by the electricity, heating, and gas networks, whose physical limits and interactions constrain the actual operational potential.

Adopting this integrated approach brings challenges in modeling, handling non-convex optimization problems, ensuring scalability, managing uncertainty, and coordinating co-optimization of interconnected networks~\cite{7842813}. To address these challenges, operation optimization becomes pivotal, enhancing reliability and economic efficiency. It supports renewable integration by determining optimal scheduling and control setpoints across multiple energy carriers over short- to mid-term horizons and sometimes day-ahead planning. Traditional scheduling approaches, such as optimal power flow, unit commitment, and economic dispatch, must be redefined to capture interactions and flexibility across interconnected networks~\cite{Malley2020}. Advanced optimization-based control, including model predictive control (MPC), further enables dynamic responses to changing conditions, enforcing constraints and maintaining desired performance through real-time control.

This review highlights the growing interest in operational optimization of IES and the recognition of the critical role that network behavior and constraints play in ensuring realistic, reliable, and efficient system operation. Although research in this timely and important field has progressed considerably in recent years, a comprehensive, structured review is needed to explore emerging techniques, highlight key trends, and identify open problems. 

\subsubsection{The Critical Role of the Network Component.}
\begin{figure}[t]
\centering
\includegraphics[width=0.5\textwidth]{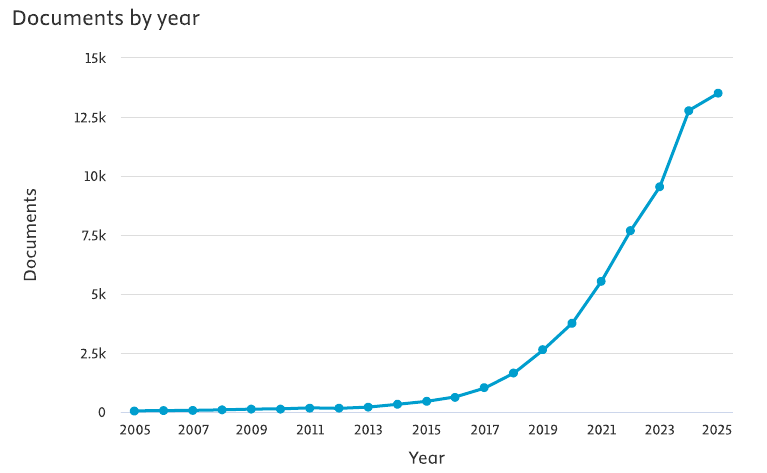}
\caption{Publication trend related to integrated/multi‑energy systems (2005–2025). Source: Scopus, retrieved [21/02/2026]. Query:
TITLE-ABS-KEY( "integrated energy system*" OR "multi* energy system*" OR "multi* carrier energy system*" OR "sector coupling" OR "energy hub*" OR "integrated electricity heat gas" OR "electric* heat gas network*" OR "multienergy system*" OR "multi-energy network*" ) AND PUBYEAR > 2004 AND PUBYEAR < 2026 AND ( LIMIT-TO (SUBJAREA, "ENGI") OR LIMIT-TO (SUBJAREA, "ENER") ).}
\label{MENscopus}
\end{figure}

To situate this review within the broader literature, we issued a Scopus query over standard integrated/multi-energy systems terminology (2005--2025), as shown in~Figure~\ref{MENscopus}. The query returns 47{,}079 documents and confirms a rapid expansion of related journal publications, especially over the last decade. This growth is particularly pronounced in the last four years, reflecting the rising recognition of the importance of these energy systems. 

Within the papers cited in this review, only a subset qualifies as \emph{network-aware}, namely, studies whose operational decisions explicitly take network topology, behavior, and constraints into account. Consequently, only a very small share of the retrieved documents is network-aware. 

This disparity highlights a critical gap between theoretical models and operational reality, which is notable given that including network behavior is essential for realistic and operationally viable solutions, even though it significantly increases the complexity of the problem.

This gap motivates the focus of this review on \emph{topology- and constraint-explicit} formulations for \emph{operational optimization and control across IES}, providing clear criteria for when they are required, conditions under which relaxations remain valid, and guidance for practical deployment.

\subsection{Related Works and Contributions}\label{subsec:contrib-positioning}

We target the \emph{minutes–hours} operational layer, where spatio‑temporal, network aware decisions materially depend on carrier topology and physics/flow constraints. We do not consider studies that coordinate only via hub‑level exchanges without carrier‑network behavior/topology, and works on sub‑second electromechanical dynamics or frequency services, which are dominated by electromechanical dynamics in the electrical grid alone; conversely, long term planning can legitimately rely on coarser or even no network models (or on the same formulations but heavily simplified). Focusing on minutes–hours maximizes the relevance of spatial feasibility and time‑coupled behavior across carriers, where realistic multi-energy operation demands explicitly accounting for the network behavior and topology, and where our synthesis adds the most value. Ignoring network physics at this layer can likely lead to infeasible dispatch (e.g., thermal advection delays or linepack constraints are violated).

\subsubsection{Related Surveys}
Recent surveys and tutorials increasingly emphasize cross‑carrier operation. A network‑analytics perspective advances standardized formulations for electricity–heat–gas (e.g., energy‑hub matrices and generalized circuit or multi‑port equivalents) and shows how these formulations provide a foundation for modeling choices in both planning and operational contexts~\cite{Huang2020}. 
A high‑level overview maps multicarrier technologies, sector couplings, and market/operational challenges for a highly electrified, renewables‑dominated future~\cite{Malley2020}. Flexibility tutorials formalize multienergy‑node envelopes and aggregation, clarifying how cross‑vector interactions can be exploited, without providing and analyzing network‑explicit operational models~\cite{Chicco2020}. 
A dynamics‑focused review highlights the shift from static to explicit dynamics across devices, district heating, and gas, stressing the role of slow thermal/gas inertia and multi‑timescale effects for operational feasibility at minute‑to‑hour horizons~\cite{Jin2024ProspectiveIESDynamics}.A taxonomy and tool survey emphasizes available software platforms rather than the role of networks~\cite{Song2022}, while whole‑energy evaluation frameworks remain conceptual and planning‑oriented~\cite{Berjawi2021}. 
The survey~\cite{Ringkjob2018_modelling_tools} catalogs simulators and long‑horizon planning frameworks without considering the network‑explicit behavior that governs operations‑time feasibility across IES. 
This review goes beyond tool taxonomies and hub‑centric surveys by setting out \emph{network‑explicit} formulations, pinpointing where simplified surrogates fail, and linking model choice to \emph{certified convex relaxations} and \emph{closed‑loop controllers} that keep the real distribution‑level IES feasible. Unlike prior reviews, it provides a structured synthesis of modeling and optimization methods that explicitly incorporate network-level constraints and interactions, and identifies actionable research directions to address scalability, uncertainty, and real-time feasibility. It highlights key findings that can help improve the design and control of systems that manage different energy types together. 

\subsubsection{Contributions}
To the best of our knowledge, this is the first review that comprehensively addresses network-aware optimization and control of IES, with emphasis on scalability, uncertainty, and formal guarantees.
This review fills that gap in the literature by
\begin{itemize}
\item surveying network-aware modeling across multiple energy networks, clarifying when network physics and topology must be modeled explicitly and when common relaxations remain valid;
\item synthesizing modeling, optimization, and control approaches specifically at the minutes–hours operational scale, highlighting where existing formulations become infeasible or lose physical accuracy;
\item providing the first integrated perspective connecting network-aware MPC, distributed optimization, and machine-learning-based methods, identifying where guarantees (feasibility, stability, convergence) are missing;
\item outlining a research agenda for scalable, physics-consistent, and uncertainty-aware operation of future multi-energy system.
\end{itemize}

\subsubsection{Outline of the Paper}
The remainder of this paper is structured as follows.
Section~\ref{Fundamentals} reviews the fundamentals of Integrated Energy Systems (IES), introducing their key components, energy carriers, coupling technologies, and typical layouts.
Section~\ref{MEN_model} presents modeling principles for IES, covering Distributed Energy Resources (DERs), dynamic components such as storage systems and buildings, and detailed modeling of electricity, heating, and gas networks, culminating in system‑level formulations.
Section~\ref{ConOpt}  surveys operation optimization and control approaches, including single‑period and multi‑period formulations, modeling and architectural choices, optimization techniques, and feedback control strategies such as MPC and reinforcement learning.
Section~\ref{Challenges} outlines open challenges and future research directions, emphasizing the need for physics‑aware modeling, scalable optimization, network‑informed learning, interoperability, and uncertainty‑aware operation.
Section~\ref{Conclusions} concludes the paper.

\section{Fundamentals of Integrated Energy Systems}~\label{Fundamentals}
An IES is a complex system, typically at the distribution level, that coordinates multiple energy vectors to deliver a balanced and efficient energy supply by leveraging synergies among electricity, gas, heating, and cooling networks~\cite{Xu2024,7842813}. Key IES components include Distributed Energy Resources (DERs), which are small- to medium-scale energy systems located near consumption points, typically on the distribution side of the electrical network~\cite{Xu2024,Chicco2020}. DERs encompass energy conversion, storage, consumption devices, and coupling elements. Examples include distributed generators (biomass units, microturbines), flexible demands (smart appliances, heating, ventilation, and air conditioning (HVAC)), and storage systems, all of which enhance local generation, flexibility, and reliability. Coupling components, such as power-to-heat (electric boilers, heat pumps (HPs), electric chillers), gas-to-power (microturbines, gas turbines), combined heat and power (CHP) units, gas boilers, and absorption chillers, convert and link energy carriers, facilitating renewable integration and decarbonization. Buildings  play a key role in IES, as they have diverse energy needs and connect different energy vectors at nodes within these networks.

Figure~\ref{MEN} illustrates an IES with three interconnected networks: electric power (solid blue lines), district heating (dashed red), and gas (dashed green). 
\begin{figure*}[t]
\centering
\includegraphics[width=\textwidth]{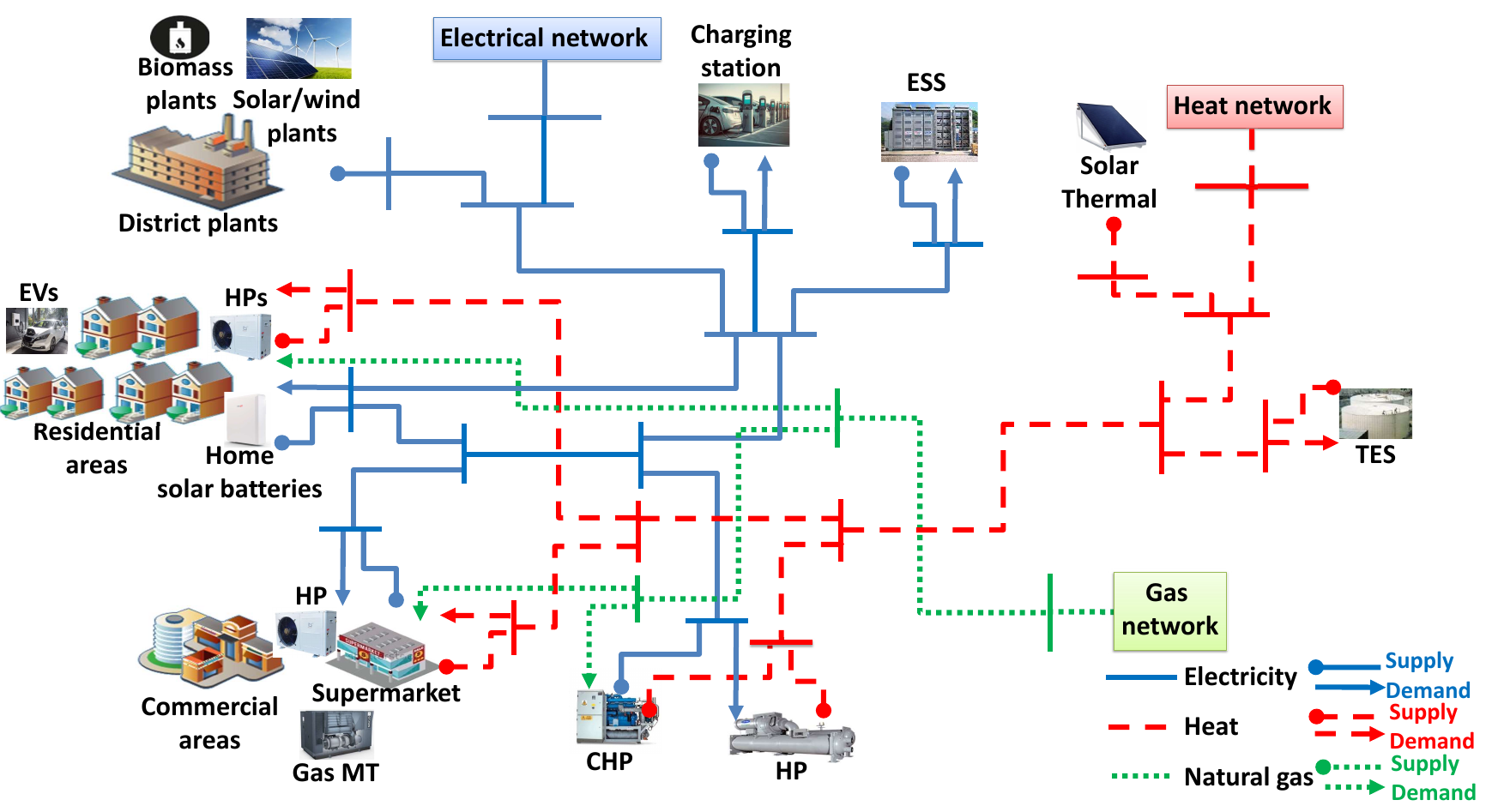}
\caption{Integrated Energy Systems conceptual architecture.}
\label{MEN}
\end{figure*}

These networks integrate diverse resources and devices. Renewable Energy Sources (RES) feed electricity into the grid, biomass supplies electricity and heat, and storage systems, such as thermal energy storages (TES), manage variability and demand-response flexibility. Electric vehicles (EVs) and charging stations provide both demand and storage capabilities. HPs, microturbines (MTs), and CHP plants convert energy forms, coordinating the networks to improve efficiency, flexibility, and resilience. Byproducts like waste heat can supply district heating, while excess electricity can be stored or converted to heat.

Emerging technologies, such as hydrogen systems (electrolyzers, fuel cells) and renewable thermal sources, offer long-duration storage, flexibility, and low-carbon heat~\cite{Xu2024}. Additional networks, such as transportation, hydrogen, and water distribution, can also be integrated; see~\cite{cascetta2009, ball2009, larry2000} for details.
\\
The subsequent section will discuss the essential principles of IES modeling necessary for developing control and optimization frameworks.

\begin{remark}
In this review paper, \textit{steady-state} refers to algebraic models that describe system behavior after all transient effects have settled, representing equilibrium conditions for given inputs. When steady-state models are applied over changing inputs, outputs vary with time, but transient dynamics are ignored since the system settles faster than the considered time scale. In contrast, \textit{dynamic models} explicitly represent time evolution through differential or difference equations, capturing transient and changes in system states. 
\end{remark}

\section{Modeling Principles for Control and Optimization of IES}\label{MEN_model}

The analyzed studies encompass a variety of system layouts for IES, ranging from 
networked buildings or community systems that include building thermal dynamics for local flexibility to multi-region IES. Despite their differences, these approaches share a common principle: integrating subsystems to exploit synergies across energy carriers, with the network component playing a central role in enabling cross-vector energy exchange and coordinated system operation. 
\\
The following subsections present an overview of the modeling approaches used in the reviewed studies, beginning with the modeling of DERs and their aggregation through the energy hub paradigm~\cite{Ding2022_energy_hub_optimization}, followed by the modeling of energy networks across various carriers.

\subsection{Modeling of DERs}
Most DERs, such as CHP units and HPs, are typically modeled using steady-state representations based on constant efficiency factors, i.e., $E_t^{\text{out}} = \eta \, E_t^{\text{in}}$, where $E_t^{\text{out}}$ is the output energy of a DER, $E_t^{\text{in}}$ is its input energy, $\eta$ is the conversion efficiency. While $\eta$ is often assumed constant for simplicity, more accurate models consider its dependence on operating conditions. In such cases, the nonlinear efficiency curve $\eta(\cdot)$ is frequently approximated by piecewise linear functions to obtain tractable optimization frameworks, involving auxiliary continuous and binary variables. The CHP operating region is sometimes represented as a polytope or as a convex combination of its extreme points to capture feasible heat and power output combinations (e.g.,~\cite{9437972,Cao2018,Zhai2020,Huang2023}). 

These steady-state models are widely accepted because the internal dynamics of many DERs typically stabilize quickly enough. However, the startup and shutdown processes of certain units, such as fuel cells, might still need to be modeled. Operational constraints such as minimum up- and down-times, ramping limits, and minimum stable output levels are also commonly included to avoid unrealistic cycling and to reflect equipment limitations in practice. In contrast, temporal dynamics are critical for components like energy storage systems and building Heating, Ventilation and Air Conditioning (HVAC) systems, which exhibit state-of-charge dynamics or significant thermal inertia that must be explicitly modeled to capture their behavior over time, as shown in~\ref{DYN}.
Some DERs, like reversible heat pumps, operate in distinct modes (e.g., heating or cooling) that cannot happen simultaneously, or that are subject to logical conditions. To model this accurately, binary variables are used to ensure that the model captures the device’s behavior realistically. Furthermore, binary variables are commonly used to represent the on/off status of distributed generators. As a result, most optimization frameworks proposed in the literature are formulated as mixed-integer programs (MIP). 
The reader is referred to~\cite{Parisio2014MPCMicrogrid,10378966,9437972} and references therein for further details on the DER modeling. 

\subsubsection{Energy Hub Aggregation}~\label{EH}
The~\textit{energy hub} is a widely adopted modeling paradigm that aggregates multiple DERs into a single modeling entity, which converts, stores, and distributes multiple energy forms~\cite{Chicco2020}.

This abstraction does not explicitly model the underlying network; instead, it treats the hub as a nodal representation. While the classical energy hub model is generally formulated under steady-state conditions, the model is often extended to a dynamic, multi-period framework when storage components, such as batteries or thermal storage, are included~\cite{Ayele2018}. 

This approach is frequently adopted in the literature to represent aggregations of devices that couple different energy vectors. For instance, the clusters of buildings and DERs shown in Figure~\ref{MEN} can be interpreted and modeled as an energy hub. Some studies model IES at the district or distribution level using energy hubs, essentially representing local distribution systems as energy hubs, while modeling the networks connecting these hubs at the transmission level~\cite{Gan2021}.

A widely used representation of an energy hub employs a single coupling matrix to describe the steady-state relationship between energy inputs and outputs, i.e., $\mathbf{y}_t = \mathbf{C} \mathbf{u}_t$, where $\mathbf{u}_t \in \mathbb{R}^{n_u}$ is the vector of input energy flows (e.g., electricity, gas, heat) at time $t$, $\mathbf{y}_t \in \mathbb{R}^{n_y}$ is the vector of output energy demands (e.g., electricity demand, heating load), and $\mathbf{C} \in \mathbb{R}^{n_y \times n_u}$ is the \emph{coupling matrix}. 
An example of a simple energy hub, comprising one CHP, one heat pump (HP), and one microturbine (MT), is shown in Figure~\ref{EH}, along with its input and output vectors and the coupling matrix, containing the electrical and thermal efficiency factors of the CHP, $\eta_e$ and $\eta_{th}$ respectively, the efficiency factor of the MT, $\eta$, and the coefficient of performance (COP) of the HP.

This compact formulation is commonly used in optimization frameworks that focus on the aggregated behavior of energy conversion and distribution without modeling internal network and device-level dynamics.

\begin{figure}[h!]
\centering
\includegraphics[width=0.45\textwidth]{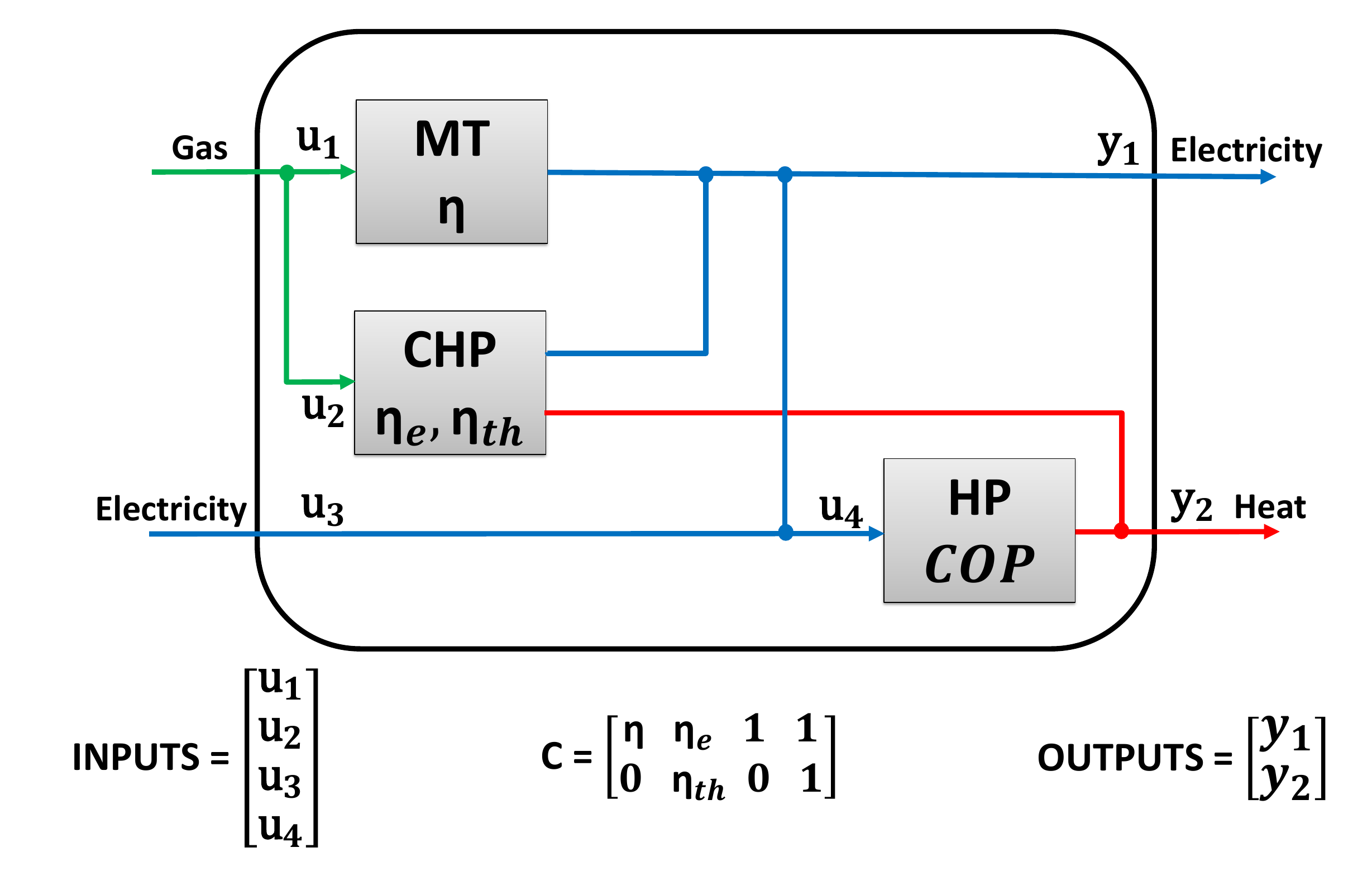}
\caption{An example of an energy hub.}
\label{EH}
\end{figure}

\subsection{Modeling of Dynamic Components}\label{DYN}
Dynamic components introduce temporal coupling into the system by linking decisions across time steps. In this section, we focus on two representative sources of flexibility: electrochemical storage systems and building-based flexible demand. Both components exhibit internal dynamics that can be captured using state-space models, enabling their integration into a multi-energy network optimization framework.
\subsubsection{Storage}
Electric storage systems provide flexibility but exhibit losses upon charging and discharging due to internal resistance. A battery system is typically modeled by using the following state-space model~\cite{10378966}. 

\begin{equation}
\begin{aligned}
x(t+1) &= (1-\alpha)x(t) 
 + \left(\eta_{\text{chg}} - \frac{1}{\eta_{\text{dchg}}}\right) z(t)\Delta t \\
&\quad + \frac{1}{\eta_{\text{dchg}}} u(t)\Delta t \\
y(t) &= x(t)
\end{aligned}
\label{eqBatt}
\end{equation}
with the introduction of the binary variable $\delta(t)$, i.e. 
 $\delta(t)=1$ for charging, $\delta(t)=0$ for discharging, and the following logical conditions
\begin{equation}\label{eqBattLog}
    u(t) \geq 0 \iff \delta(t) = 1 \, (\text{charging}), 
\end{equation}
where $x(t)$ is the amount of energy stored in the battery, $u(t)$ is the control input, i.e., charging/discharging power input, $\eta_{\text{chg}}$ and $\eta_{\text{dchg}}$ are the charging and discharging efficiencies, respectively, and  $\alpha$ is the battery self-discharge.
As shown in~\cite{BEMPORAD1999407,10378966}, the logical condition~\ref{eqBattLog} and the product $\delta(t) u(t)$ can be equivalently expressed using mixed-integer linear inequalities and the auxiliary continuous variable $z(t) = \delta(t) u(t)$. 

An alternative to the use of binary variables is to impose a complementarity constraint that forces the product of the charging and discharging power variables to be zero. While this avoids explicit mode‑selection binaries, the resulting constraint is non-convex and therefore also challenging to handle in practice~\cite{Jia2024}. Models that exclude both binary variables and complementarity constraints do not explicitly prevent simultaneous charging and discharging, leading to physically inaccurate behavior (e.g.,~\cite{ALIZADEH2024122162}). 
\subsubsection{Buildings and Flexible Demand}
HVAC systems are the primary source of flexible demand, leveraging building thermal inertia as a form of virtual thermal storage. Building thermal dynamics can be modeled with varying levels of detail, depending on the application and the desired accuracy.
Some models represent buildings only through their electrical and thermal loads, taking into account thermal comfort and the Predicted Mean Vote (PMV) indicator, which quantifies occupant thermal comfort, but without including dynamic behavior~\cite{Li2022}.
A more refined and typical approach involves first-order equivalent thermal models that describe indoor temperature dynamics over time, averaged over a sampling period. These typically linear models use heating power to the building or zone as input and are often based on the RC (resistor-capacitor) analogy~\cite{Lu2023}. This analogy models buildings and their HVAC systems as dynamical systems by drawing a parallel between heat transfer and electrical current flow: temperature is analogous to voltage, and heat flow to current, with thermal resistances and capacitances representing the building's thermal properties. Consequently, temperature dynamics in walls and indoor zones are described by differential equations, which are discretized into difference equations to capture their evolution over discrete time steps. An illustrative example is:
\[
T(t+1) = T(t) + \frac{\Delta t}{RC} \left( T^o(t) - T(t) \right) + \frac{\Delta t}{C} Q(t),
\]

where \( T(t) \) is the (average) indoor temperature at time step \( t \), \( T^o(t) \) is the outdoor temperature at time step \( t \), \( Q(t) \) is the heat input (e.g., from HVAC) at time step \( t \), \( R \) is the thermal resistance, \( C \) is the thermal capacitance, and \( \Delta t \) is the sampling time. 
More accurate models represent the thermal zone with separate nodes for walls and room air, connected through thermal resistances and capacitances, and model heat exchange using linear difference equations for each wall~\cite{Jia2024}. Further detail is added by explicitly modeling the inputs in terms of mass flow rates for the building's heat load and the temperatures of the supply and return nodes in the thermal network. However, to maintain linearity and reduce complexity by avoiding bilinear terms, mass flow rates are typically treated as known parameters, which reduces the ability to fully leverage the system’s thermal flexibility~\cite{Tan2020}. Other studies emphasize that the building’s thermal inertia can act as virtual thermal storage and be incorporated into optimization frameworks to better represent the building's flexibility and enable more effective energy management~\cite{Cesena2020}. 

When buildings incorporate local generation and storage, they can be represented as IES themselves, which include DERs like batteries, thermal storage tanks, heat pumps, and flexible loads. To capture the complex interactions among these components, the Mixed Logical Dynamical (MLD) framework is often employed. This hybrid modeling approach integrates continuous dynamics (e.g., temperatures and storage energy levels), discrete logic (e.g., on/off states, operating modes), and operational constraints into a unified mathematical structure. The MLD model includes state variables representing indoor temperatures and storage levels, continuous control inputs such as power and temperature setpoints for DERs and HVAC systems, and binary variables that express logical decisions like device status or mode changes. Nonlinear inequalities can be included in order to more accurately represent the nonlinear behavior of DERs, such as nonlinear COPs and efficiencies~\cite{10378966}.

\subsection{Modeling of Energy Networks}
In this section we summarize network modeling choices for IES with emphasis on operational time scales (minutes–hours), highlighting the modeling assumptions that preserve physical feasibility while remaining tractable for control and optimization.

The network layout and topology are commonly captured through a graph-like structure, where a directed graph $\mathcal{G} = (\mathcal{N}_n, \mathcal{E}_n)$ represents a given network of energy type $n$. The set of nodes, $\mathcal{N}_n$ represents buses, substations, or junction points where energy is injected, withdrawn, or routed. The set of edges $\mathcal{E}_n \subseteq \mathcal{N}_n \times \mathcal{N}_n$ corresponds to the physical infrastructure, such as power lines, gas pipelines, or heat distribution pipes, that enables energy transfer between nodes.
Modeling integrated energy networks requires accounting for distinct temporal dynamics and spatial scales. Electrical networks operate on fast timescales (sub-seconds to tens of seconds), while thermal and gas networks evolve more slowly (minutes to hours). Consequently, electrical networks are generally assumed to operate in steady-state for IES optimization. Most studies adopt a discrete-time modeling approach with hourly sampling, though finer resolutions (15–30 minutes) are used in~\cite{10378966,9437972,10252152,Jin2021}. Exceptions include~\cite{Han2024}, which employs continuous-time dynamics with modified droop control for frequency stability, and~\cite{Jiang2021}, which uses 5-minute intervals for robust scheduling. MPC is applied in~\cite{Dou2020} to refine hourly dispatch with minute-level updates.

The spatial scale of IES primarily spans the distribution level, including urban districts~\cite{10378966,9437972,Haghighi2015,Xu2020}, multi-energy microgrids~\cite{Zhang2024,Xiao2023,9596598}, and community or park-level systems~\cite{Lu2021,Li2022,Ji2024,Wu2023a}. These systems integrate electricity, heat, gas, and sometimes hydrogen~\cite{Li2025,Han2023} via local energy hubs~\cite{Huo2019,Gan2024}. While most work targets local or regional systems, some studies address transmission-level modeling for multi-region IES, transmission congestion, or the integration of large-scale renewables~\cite{ZHANG2024123593,Han2023,Wang2019,Chen2023b}. Applications also include ports and transport networks integrating EVs and logistics equipment through nodal energy hubs~\cite{Chen2023a,Cesena2019}. Several frameworks model building-level flexibility and virtual storage using thermal inertia~\cite{Wu2023a,Lu2021,Cesena2020}, emphasizing the role of buildings in district energy systems. A few works explicitly address the interaction between transmission and distribution systems, especially when coordinating heat and power through CHPs and heat pumps (e.g.,~\cite{Deng2020,Zhang2023}). 

\subsection{Modeling of Energy Networks in IES}~\label{Network_model}
The following sections provide insights into the modeling of each energy network. For each carrier (electricity, heating, gas), we outline the core equations/assumptions used at operational horizons and indicate where common relaxations and approximations (e.g., second-order cone relaxation or piecewise linearizations) remain adequate and where they risk infeasibility or loss of accuracy.

\subsubsection{Electrical network}

\paragraph{Operational scope and time scales}
Given the typical operational time scales of interest, on the order of minutes to an hour, steady-state power flow models are sufficient for optimization and control tasks. Fast electrical dynamics, such as generator frequency response and inertial effects, are not relevant unless the study explicitly targets stability or frequency control. However, certain control-oriented applications may incorporate simplified dynamics to account for effects such as inverter-based voltage regulation or load switching.
\paragraph{Core formulations}
Electrical distribution systems are typically represented using branch flow formulations (see Box 1) or their DistFlow approximations, which assume a balanced three-phase radial system and negligible shunt admittances. At the transmission level, the linear DC power flow model is commonly adopted, relying on key assumptions such as negligible line resistance and reactive power flows—assumptions that generally hold true in high-voltage transmission networks~\cite{Chen2023b}. DC power flow approximations focus on active power flows, phase angles, and reserve constraints, while often neglecting reactive power and voltage magnitude variations.
\paragraph{Relaxations, approximations, and modeling extensions}
Due to the non-convex nature of the power flow equations, convex relaxations are commonly applied to facilitate efficient solution of optimization problems. One widely used approach is the second-order cone programming (SOCP) relaxation of the branch flow model. Under assumptions such as radial topology, negligible line losses, and flat voltage deviation~\cite{Zhao2021}, the non-convex equality constraints of the power flow problem are commonly relaxed into convex inequalities. SOCP formulations have been shown to yield exact or nearly exact solutions in several practical cases~\cite{Gan2015}. These relaxations handle voltage, active/reactive power flows, and line capacity constraints with improved computational tractability. Some models include device-specific constraints like tap changers, voltage/VAR regulation equipment, and inverter output limits. Energy storage systems are typically modeled dynamically, with state-of-charge (SoC) variables included. In some works, linearized versions of the DistFlow model are adopted~\cite{Hu2024}.

\subsubsection{Heating network}
\paragraph{Operational scope and time scales}
Heating and cooling networks are typically modeled using either steady-state, quasi-dynamic, or, less commonly, fully dynamic formulations (see Box 1). Cooling networks are rarely included~\cite{10378966,9437972}, with most studies focusing primarily on heating systems. Heating networks typically operate under two modes: CF-VT (constant flow–variable temperature), where mass flow remains constant and supply temperature is adjusted, and VF-CT (variable flow–constant temperature)~\cite{Lu2021}, where temperature is fixed and flow rate varies. In much of the literature, CF-VT and VF-CT modes are treated interchangeably with quality and quantity regulation, respectively~\cite{Huang2023,Guo2022,Lu2021,Jin2021}. The CF-VT mode is commonly used in optimization studies due to its computational simplicity and its widespread use in practice~\cite{Zhou2020,ALIZADEH2024122162}, decoupling the hydraulic and thermal models by assuming constant mass flow rates and allowing temperature to vary. However, quantity regulation (VF-CT) is increasingly adopted in modern systems for its greater operational flexibility and potential efficiency gains, particularly when supported by variable-speed pumping and advanced control strategies~\cite{Jin2021,Hu2024}. Nonetheless, fixing either flow or temperature helps simplify computations but limits the control flexibility and potential benefits achievable in real-world systems.
\paragraph{Core formulations}
To reduce the complexity of the equations governing the networks, several assumptions are widely adopted, such as treating the fluid (typically water) as incompressible with constant density; assuming constant friction factors in pipes; and treating warm and cool pipe networks as identical and symmetric, with fixed return temperatures at substations~\cite{Ayele2018,10378966}. In many cases, conduction is neglected. Another underlying assumption is that mixing occurs faster than thermal transport in the pipes. The momentum equation is often approximated or replaced by algebraic pressure-drop relationships (such as Darcy–Weisbach or empirical formulas), because transient pressure-wave propagation effects are usually negligible in low-speed incompressible flow. These commonly adopted assumptions are generally realistic for heating networks.
\paragraph{Relaxations, approximations, and modeling extensions}
In the operational optimization of heating networks, steady-state models are widely used for their computational tractability. 
However, steady-state equations are not fully adequate at typical operational time scales, as they neglect essential thermal dynamics that influence control actions and thermal comfort—most notably, the transport delay of thermal energy in pipelines. To approximate these effects without modeling full dynamics, some studies adopt steady-state formulations that incorporate heat losses through exponential functions. For more accurate representations of delay and transient behavior, dynamic models are required, which are governed by partial differential equations (PDEs). To reduce complexity, such models are often simplified using spatial discretization methods, such as finite-volume or finite-difference schemes~\cite{Huang2023,10252152}, or replaced with simplified dynamic formulations such as lumped-parameter ordinary differential equations (ODEs)~\cite{Guo2022,Ji2024}. 
These methods retain essential transport-delay effects while remaining suitable for control and optimization applications. Heating networks may include components such as heat exchanger stations, thermal energy storage, and distributed heat sources (e.g., CHP units, heat pumps), with models accounting for supply and return temperatures, nodal temperatures, inlet–outlet relationships, and pipe heat losses. Despite the simplifying assumptions applied to the physical equations, further relaxations and approximations are often required to ensure computational tractability. To handle the non-convex nature of the heating network equations, convex relaxations and linear approximations are frequently employed~\cite{Liu2019,Chen2023,Chen2023a}. Nodes are modeled with energy conservation and mixing equations, while pipeline heat transfer is commonly linearized or approximated using Taylor expansions. Temperature-dependent bilinearities are addressed using McCormick envelopes, piecewise linear approximations, or first-order Taylor expansions~\cite{Zhang2023,Ji2024}. Convex relaxations such as SCOP have been applied to pipeline temperature-drop equations, while nonlinear hydraulic relationships, including pump power consumption and pressure loss, are often approximated by linear or piecewise linear functions~\cite{9437972,Chen2023a,10378966}.

\subsubsection{Gas network}
\paragraph{Operational scope and time scales}
Gas networks are usually modeled with steady-state flow and pressure relationships under isothermal conditions, neglecting energy conservation (see Box 1). These steady-state models capture mass conservation and pressure–flow relationships under the assumption of isothermal flow, thereby neglecting the energy conservation equation. Such simplifications are generally justified for the time scales of interest, especially in small or moderately sized networks where pressure dynamics, transient effects, and linepack dynamics are often negligible~\cite{Wang2018,Chen2023b}. However, in larger networks, dynamic models governed by PDEs are essential to capture pressure transients and linepack dynamics. These PDEs can typically be solved via finite-volume or finite-difference methods to ensure tractability, though they remain computationally demanding. To make them tractable for practical applications, approximated dynamic characteristics are used, which reduce computational cost but can compromise the accuracy of dynamic behavior~\cite{Zhao2021}. In~\cite{Wang2024_DynamicEquivalentGasStorage}, the gas network is represented as a dynamic multi‑port storage model at the interfaces with gas turbines and power‑to‑gas units; this retains linepack and time‑coupling and avoids solving the PDEs explicitly.
\paragraph{Core formulations}
Compressors are sometimes modeled with simplified flow constraints or represented by a compression ratio, although the energy consumption of compressors is not often included~\cite{Haghighi2015,Lv2022}. For gas–power coupling with electric‑driven compressors, a reduced‑order, transfer‑function formulation with a two‑port compressor model provides analytical operating envelopes that can be enforced directly in optimization, avoiding full transient partial‑differential‑equation simulations while retaining the relevant dynamics~\cite{Huang2025_TSG_EDC_DynamicModeling}.

A few models consider gas quality indices such as specific gravity or calorific value, or the interaction between gas and power flows, especially in systems involving power-to-gas (P2G) or gas-fired generators~\cite{Gan2021,Zhao2021}. When composition varies (for example, due to power‑to‑gas injections or hydrogen/biogas blending), expressing nodal demands in energy units and including gas‑quality tracking helps prevent situations where a lower gross calorific value meets volume targets yet under‑delivers energy or does not meet minimum‑quality requirements at some nodes~\cite{Osiadacz2020_GasDistributionMES}.
\paragraph{Relaxations, approximations, and modeling extensions}
In general, the governing equations remain nonlinear and non-convex, particularly due to the Weymouth equation, which defines a nonlinear relationship between gas flow and pressure drop. To address this, several relaxation and approximation techniques have been developed. Second-order cone programming (SOCP) relaxations provide a convex surrogate for the Weymouth equation and are widely adopted in both power and gas network optimization problems~\cite{ALIZADEH2024122162,Xie2020,Cesena2019}. Other methods include linearized gas flow equations that approximate nonlinearities under flat pressure assumptions, and McCormick envelopes for bilinear terms~\cite{Zhao2021a,Zhang2024}. Piecewise linear approximations and first-order Taylor expansions are employed to linearize nonlinear constraints such as pressure drop~\cite{Lyu2023,Gao2022,Liu2019}.

\begin{figure*}[!t]              
  \centering
    \begin{tcolorbox}[
  enhanced,
  width=\textwidth,
  title=Box 1: Main Equations for Network Modeling, 
  colback=lightblue, colframe=blue!40!black,
  colbacktitle=lightblue_title, coltitle=black,
  fonttitle=\bfseries, boxrule=0.5pt, arc=4pt,
  left=6pt, right=6pt, top=6pt, bottom=6pt,
  breakable=false              
]
\paragraph{Electric network} 
\textbf{Branch flow equations.} The branch flow model describes power flows using complex voltages and currents. For each node \(i\) and \(j\), the model consists of:
$V_i - V_j = z_{ij} I_{ij}, \quad
 S_{ij} = V_i I_{ij}^* = P_{ij} + j Q_{ij}, \quad 
\sum_{k \in \mathcal{C}(j)} S_{jk}
\;=\;
s_j
\;+\;
\sum_{l \in \mathcal{P}(j)} \Big( S_{lj} - z_{lj}\,|I_{lj}|^2 \Big),$
where \(V_i, V_j\) are complex node voltages, \(I_{ij}\) is the complex current on the branch, \(z_{ij} = r_{ij} + j x_{ij}\) is the line impedance, \(S_{ij}\) is the complex power flow from \(i\) to \(j\), and \(s_j\) is the \emph{net complex power injection at node \(j\)} (generation positive; load negative). We take \(\mathcal{C}(j)=\{k:(j,k)\text{ is a child edge}\}\) and \(\mathcal{P}(j)=\{l:(l,j)\text{ is the (unique) parent edge in a radial network}\}\);
\paragraph{Heating network} 
\textbf{Mass conservation equations.}
Mass conservation ensures that the mass flow rate entering a node equals the rate leaving it:$\sum_{j \in \mathcal{N}_i^\text{in}} \dot{m}_{ji} = \sum_{k \in \mathcal{N}_i^\text{out}} \dot{m}_{ik}$, where \(\dot{m}_{ij}\) is the mass flow from node \(i\) to \(j\). \\
\textbf{Pressure drops equations.}
Pressure drops along the pipe are typically modeled using Darcy-Weisbach or similar empirical laws: $\Delta p_{ij} = K_{ij} \dot{m}_{ij} |\dot{m}_{ij}|$, where \(K_{ij}\) is a resistance coefficient dependent on pipe geometry and fluid properties.\\
\textbf{Energy conservation at nodes.}
The temperature mixing at nodes is given by: $T_n = \frac{\sum_i \dot{m}_i T_i}{\sum_i \dot{m}_i}$, where \(T_n\) is the node temperature, \(\dot{m}_i\) and \(T_i\) are the mass flow rate and temperature of the \(i\)-th incoming pipe.\\
\textbf{Temperature drops equations.}
Temperature loss due to heat exchange with surroundings is given by: $T_j = T_i e^{-U_{ij} L_{ij} / (\dot{m}_{ij} c_p)} + T_\text{a} \left(1 - e^{-U_{ij} L_{ij} / (\dot{m}_{ij} c_p)}\right)$,
where \(U_{ij}\) is the heat transfer coefficient, \(L_{ij}\) is pipe length, and \(T_\text{a}\) is ambient temperature.\\
\textbf{Dynamic Energy Conservation in Pipes (with Losses).}
The temperature evolution along a pipe is given by: $\rho c_p A \frac{\partial T(x,t)}{\partial t} + \dot{m} c_p \frac{\partial T(x,t)}{\partial x} = -k_{\text{loss}} \left(T(x,t) - T_{\text{a}}\right)$, where \( \rho \) is the fluid density, \( c_p \) the specific heat, \( A \) the cross-sectional area of the pipe, \( \dot{m} \) the mass flow rate, \( T(x,t) \) the fluid temperature, \( T_{\text{a}} \) the ambient temperature, and \( k_{\text{loss}} \) the linear heat loss coefficient.
\paragraph{Gas network} 
\textbf{Pressure drops equations.} 
The Weymouth equations are commonly used for pressure drops: $p_i^2 - p_j^2 = K_{ij} \dot{m}_{ij} |\dot{m}_{ij}|$, where \(p_i\), \(p_j\) are pressures at the ends of the pipelines, and \(K_{ij}\) is a pipeline-dependent constant.\\
\textbf{Mass conservation equations.}
The conservation of mass is described as: $\sum_{j \in \mathcal{N}_i^\text{in}} \dot{m}_{ji} = \sum_{k \in \mathcal{N}_i^\text{out}} \dot{m}_{ik}$, similar in structure to heating networks.\\
\textbf{Dynamic gas flow in pipelines.} 
The transient behavior of gas flow is governed by the following equations:\\
\textit{Continuity of flow (mass conservation)}:$\frac{\partial p}{\partial t}
+ \frac{RT}{A}\,\frac{\partial \dot m}{\partial x} = 0$. Linepack dynamics are captured through the pressure time derivative. \\
\textit{Momentum conservation}: $\frac{\partial \dot m}{\partial t}
+ A\,\frac{\partial p}{\partial x}
+ \frac{f}{2D}\,\frac{\dot m \lvert \dot m \rvert}{\rho A} = 0, $
where \( p(x,t) \) is the gas pressure, \( \dot m(x,t) \) is the mass flow rate, \( A \) is the pipe cross-sectional area, \( \rho = p/(RT) \) is the gas density under the isothermal ideal-gas assumption, \( f \) is the Darcy--Weisbach friction factor, and \( D \) is the pipeline diameter.
\end{tcolorbox}
\end{figure*}

\subsection{Overall System-Level Modeling of IES}

Generally, the dynamic and steady-state behavior of a network can be represented by a system of nonlinear differential–algebraic equations based on physical principles. Dynamic models illustrate how fluid mass flows, pressures, and temperatures evolve over time (and space, when applicable), and can be discretized effectively using various methods such as the Euler method, bilinear transformation, or finite-difference methods. 
We integrate carrier‑level models with device and building dynamics into a system‑level IES representation, and highlight where hybrid/discrete decisions and nonlinear physics increase problem size and computational burden for operations‑time optimization.

The discretized dynamic model of an energy network can be written in the following general and compact form
\begin{flalign}\label{eq:dynamic}
	\begin{cases}
		\bm{x}_{\eta}(t+1) =  f_{\eta}(\bm{x}_{\eta}(t), \bm{u}_{\eta}(t) ), \\
		\bm{y}_{\eta}(t) =  g_{\eta}(\bm{x}_{\eta}(t), \bm{u}_{\eta}(t) ),
	\end{cases}
\end{flalign}
where $\eta \in \{\text{DHC}, \text{G} \}$ denotes the DHC or the gas network and $t$ the time index; a similar state-space model can be derived for a distribution network. The state vector, $\bm{x}_{\eta}$, of appropriate dimensions, includes temperatures of the gas and heat transfer fluid at the different nodes, heat energy stored at the various nodes. The input vector, $\bm{u}_{\eta}$, includes heat/cooling source power inputs, pumps/compressors power inputs, as well as mass flows and pressure increases, heat/cooling and gas power inputs at the network nodes, temperature setpoints of heat exchangers. The vector $\bm{y}_{\eta}$ consists of the outputs of interest, which can include temperatures, mass flows, energy levels.
Disturbances and measurement noises can be added as appropriate, such as the ambient temperature and the heat/cooling source power input (e.g., in case the heat source is a solar thermal power plant).

A system of nonlinear equations, of appropriate dimensions, is also needed to include essential equations, in particular power flow equations, nodal power balances, pressure drops, loop pressure and head loss equations as well as mass conservation. This can be written compactly as 
\begin{equation}\label{eq:ss}
F_{\eta}(\bm{y}_{\eta}(k), \bm{x}_{\eta}(k), \bm{u}_{\eta}(k) ) = 0.
\end{equation}
The model~\eqref{eq:dynamic}–\eqref{eq:ss} needs to integrate the models of the individual DERs and buildings connected to the nodes of the various energy networks, including additional power inputs/outputs to the nodal balance equations, building power demands, and additional states representing the building thermal dynamics and storage levels when batteries and/or thermal storage units are present. Binary variables are needed to capture bidirectional power and fluid flows and to represent the hybrid nature of DERs such as batteries and heat pumps. Integer variables may also be required to model the various operating modes of generating units like CHPs.

The model~\eqref{eq:dynamic}-\eqref{eq:ss} represents a highly complex, large-scale system that is computationally intensive and not well-suited for control design and optimization. Convexification and approximation techniques are typucally applied to make the model more tractable, which commonly necessitate the introduction of auxiliary and additional binary variables (see~\cite{10378966}, for example).
\paragraph{Operational Modeling Choices in IES}
The required level of dynamic modeling in IES depends largely on the specific focus—whether operational optimization, control, stability analysis, or planning. For operational time scales (minutes to about an hour), dynamics are included only if they meaningfully influence decision-making and system performance within that window. Certain processes, such as pipe/soil conduction in heating networks, flow inertia in incompressible water networks, thermal transients in gas pipelines, can be neglected or represented using steady-state or quasi-steady algebraic relations, while other processes, such as advection of temperature in heating pipes or mass and momentum transport along gas pipelines, should be retained as dynamic (PDE-based) equations. 
In buildings, thermal dynamics are essential, as internal temperatures evolve over tens of minutes and are commonly captured using differential or difference equations. For storage systems, dynamic models are crucial to represent state-of-charge evolution and associated time-dependent constraints. In contrast, DERs usually do not require dynamic modeling at this time scale, as their outputs can be adequately represented as algebraic functions of inputs, which may also be time-varying. Similarly, CHPs are often represented by static or piecewise algebraic power–heat relationships. 
\\
Table~\ref{Table1} summarizes the main equations and physical variables for electrical, heating, and gas networks, distinguishing between steady-state and dynamic formulations, with a primary focus on distribution-level networks. 
The table highlights which processes must be treated dynamically and which can be approximated as steady for minute-to-hour operation. These equations capture the hydraulic, thermal, and thermodynamic components of the networks, enforcing conservation of mass, momentum, and energy, while remaining physically consistent~\cite{danko2017model}.

\begin{table*}[t]
\centering
\footnotesize
\caption{Main equations in steady-state and dynamic network modeling}
\label{Table1}
\begin{adjustbox}{width=\textwidth}
\begin{tabular}{|>{\raggedright\arraybackslash}p{2.1cm}|
                >{\raggedright\arraybackslash}p{3.9cm}|
                >{\raggedright\arraybackslash}p{3.9cm}|
                >{\raggedright\arraybackslash}p{3.0cm}|}
\hline
\textbf{\mbox{Network}} &
\textbf{Steady-State Modeling} &
\textbf{Dynamic Modeling} &
\textbf{Main Variables} \\
\hline

Electrical Network &
Bus injection / branch flow &
Not considered (at operational scale) &
Active/reactive power flows, voltages, currents, phase angles \\
\hline

Heating Network &
Mass conservation at nodes; energy conservation at nodes (temperature mixing); pressure and temperature drops in pipes &
Energy conservation along pipes; thermal dynamics of nodes (building thermal mass) and storage elements &
Temperatures, heat flows, mass flow rates, pressures \\
\hline

Gas Network &
Mass conservation at nodes; pressure drops in pipelines &
Mass conservation along pipelines; momentum conservation &
Pressures, mass flow rates \\
\hline
\end{tabular}
\end{adjustbox}
\end{table*}

The corresponding mathematical formulations of these equations are outlined in Box 1~\cite{Huang2020,danko2017model,Li2025} for interested readers.

\subsection{Alternative modeling approaches}

Beyond traditional physics-based formulations, several alternative methods have been explored.

The \textit{generalized electric circuit theory} provides a unified way to represent district heating and gas networks by analogy with electrical circuits, using R/L/C‑type elements and port‑based (two‑port/multi‑port) transfer‑function models in the Laplace/frequency domain to capture temperature, pressure, and flow dynamics in a standardized form~\cite{Huang2020}. While these linear transfer‑function surrogates and idealized components can simplify the inclusion of slow thermal and gas dynamics in planning and operation models, parameter identification remains non‑trivial, and—when higher‑order models, multiple boundary ports, or broad frequency coverage are retained—mixed time‑scale effects and non‑negligible computational costs can still arise, even though the intent is to reduce the burden of full PDE models~\cite{Huang2020}. A recent variant, the Energy‑Circuit Method (ECM), recasts pipeline PDEs into frequency‑domain lumped two‑port equivalents and assembles network‑level generalized admittances per frequency component; combined with historical boundary conditions to obtain particular time‑domain solutions, the method is used to implement dynamic state estimation, dynamic energy‑flow analysis, optimal energy flow, and security assessment~\cite{Chen2022ECM}. ECM’s behavior depends on design choices (e.g., frequency selection, parameterization, and the surrogate‑initialization scheme), and the evidence presented so far is based on comparisons with the authors’ own FDM implementations and on engineering case studies; broader benchmarking—e.g., against external datasets—has not yet been explored. 

\textit{Machine learning} (ML) techniques have been applied to power grids~\cite{GONG2023121740}, district heating and cooling networks, and MES~\cite{ZHANG2023113688, ZHOU2024114466, kamper2023data}. Some works target data-driven modeling of network components~\cite{ZHOU2024114466}, while others pursue holistic system-level representations without explicit physical network modeling~\cite{kamper2023data}. Only a few studies learn or approximate network infrastructure behavior directly; among them, physics-informed approaches~\cite{ZHANG2023113688, ZHOU2024114466} integrate physical laws to improve generalizability and robustness. Despite progress, ML-based IES modeling remains limited due to challenges in capturing multi-scale dynamics and cross-vector coupling, representing nonlinear interactions, handling heterogeneous and often scarce measurement data, and ensuring interpretability, physical consistency, and robustness. Additional limitations include data scarcity in thermal and gas networks, and the difficulty of guaranteeing physical consistency and safety in learned models.

\textit{Port-Hamiltonian} (PH) methods offer a modular, energy-based, thermodynamically consistent modeling paradigm that supports control design via passivity and energy shaping. Existing PH frameworks increasingly appear in the literature but typically address single-domain systems or omit full network representations~\cite{6880409}. Maintaining passivity and energy consistency under discretization or model-order reduction typically requires structure-preserving techniques. While such approaches ensure physical interpretability and stability, they may increase implementation complexity and computational overhead, particularly for large-scale, nonlinear, or hybrid multi-energy systems. Moreover, constructing physically meaningful Hamiltonians and consistent interconnection/dissipation structures at scale can be non-trivial in multi-domain networked applications.

Overall, these approaches are promising but still face significant challenges and require further exploration.

\section{Operation Optimization Methods for Integrated Energy Systems}\label{ConOpt} 

Operation optimization aims to ensure efficient, reliable, and cost-effective IES operation across timescales from seconds to days. It can involve optimal dispatch, scheduling, and control strategies for multi-energy resources under various forms of uncertainty, such as variable renewable generation, fluctuating demand, and market volatility. Approaches include deterministic, stochastic, and robust optimization, as well as rolling-horizon and MPC strategies. Traditional paradigms struggle with IES complexity, data volume, and real-time requirements.
The following sections review network-aware studies. We begin with optimal power flow (OPF) and multi-energy flow (MEF) formulations, which determine optimal resource allocation at a single time instant subject to network constraints. These formulations are then extended to multi-period optimization frameworks to capture inter-temporal couplings such as storage dynamics, ramping limits, and thermal inertia. Such models are commonly applied in day-ahead planning, intra-day scheduling, and short-term operational control, covering time horizons from minutes to hours.

\subsection{Energy Flow Problems}
Here we review network‑aware, single-period, energy‑flow formulations, clarifying the role of convex relaxations, exactness conditions, and computational trade‑offs when coupling electricity with heating and gas networks. 
\paragraph{Classical and Convexified MEF Formulations}
Several studies address the MEF problem by optimizing resources under network constraints over short periods (30–60 min), typically using deterministic and centralized formulations. Electricity–heat coupling is often solved via Newton–Raphson without formal convergence analysis~\cite{Ayele2018}, while coordinated power–gas flow is reformulated as MILP through piecewise linearization of gas constraints~\cite{Shao2017}, and convex optimization is applied to electricity–gas systems incorporating transient gas flow via PDE discretization~\cite{Chen2018}. To improve tractability, convex relaxations such as SOCP are widely used. For the electrical distribution network, SOCP relaxations of the branch‑flow model can be exact in radial networks under suitable operating conditions (e.g., when voltage bounds are not binding)~\cite{Gan2015}. 
Convex restrictions of power flow feasibility sets identify guaranteed feasible injection subsets, though their conservativeness remains unexamined \cite{10237300}. Uniqueness of nonlinear steady-state gas flow solutions is proven for arbitrary topologies, with a provably exact mixed-integer quadratically constrained quadratic program (MI-QCQP) relaxation addressing bilinearities, is proposed in~\cite{PAN2016230}, while mixed-integer SOCP reformulations convexify integrated electricity–gas problems~\cite{Wang2023}. Convex formulations for CHP dispatch employ bound tightening and polyhedral relaxations to improve tractability \cite{Ding2022}. 
\paragraph{Physical Fidelity, Scalability, and Advanced MEF Approaches}
Physical fidelity is critical for gas systems: linepack provides short-term storage, but inaccurate modeling misrepresents flexibility, causing infeasible schedules or curtailment; steady-state models fail to capture linepack dynamics, and relaxations (MISOCP, MILP) often overestimate flexibility, making exact nonlinear formulations preferable for operational decisions~\cite{Zhu2021}. For large-scale heat–electricity systems, partitioned decoupling with proven convergence conditions supports steady-state analysis~\cite{Zhang2021}, while finite-difference discretization and iterative decomposition handle coupled thermal–hydraulic dynamics in~\cite{Yao2021,Yao2022}, and data-driven hydraulic–thermal models approximate water flow and temperature interactions in~\cite{Guo2023}. 
Computational efficiency and robust solution approaches are addressed in several studies, such as fast decoupled algorithms replacing Newton–Raphson Jacobians~\cite{Chen2020} and Levenberg–Marquardt solvers with adaptive damping for robustness~\cite{Huang2023a}. 
Dynamic formulations combine steady-state AC/DC power flow with discretized thermal models and SOCP relaxations for time-coupled decisions~\cite{Huang2023}, while probabilistic MEF for electricity–gas systems with hydrogen injection employs cumulant-based methods and Nataf transformations to capture correlated uncertainties~\cite{Zhang2022}. Machine learning approaches are emerging, including topological graph attention convolutional networks for adapting to changing topologies and uncertain RES outputs~\cite{Wu2024}, and physics-driven deep learning combined with SOCP for real-time feasible energy management~\cite{Clegg2016}. Overall, most methods for MEF are deterministic and centralized, relying on relaxations or heuristics. Literature focuses on steady-state analysis, while transient dynamics, multi-energy coupling, and correlated uncertainties remain challenging. Convex relaxations may yield infeasible solutions for the original problems, while data-driven methods require large datasets and provide convergence or accuracy primarily on an empirical basis.

\subsection{Models, Architectures, and Optimization Techniques}

In this section, we catalog the modeling/architecture choices (steady‑state vs. dynamic; centralized vs. distributed/hierarchical) and the solution methods, e.g., SOCP, mixed-integer linear/nonlinear programming (MILP/MINLP), decomposition, indicating where scalability and formal guarantees are present or lacking.

Tables~\ref{tab:model_network_architecture} and~\ref{tab:methods_uncertainty} summarise the modeling choices, network representations, decision architectures, and the optimization and control methods used for \emph{multi-period} IES operation.
In addition to these works, other studies explore resilience and distributed control. For example,~\cite{ZHANG2024123593} proposes a bi-level framework for infrastructure reinforcement;~\cite{9743661} coordinates energy hubs via distributed optimization robust to cyberattacks;~\cite{Han2024} models DERs and heat exchangers as agents in multi-agent control using virtual leaders and containment strategies.

\subsubsection{Models and Architectures}

Table~\ref{tab:model_network_architecture} summarizes the models, energy networks, and decision-making architectures adopted in the reviewed literature. Most works rely on steady-state formulations that represent system operation with algebraic constraints, while dynamic representations remain comparatively limited and typically confined to component-level models, such as batteries or thermal storage units (~\cite{Wang2024,Lv2022,Liu2022}). The majority of studies consider coupled electricity-heat systems, whereas the inclusion of gas, water, or transport networks remains less common (see~\cite{Cesena2019,ALIZADEH2024122162,Xu2020,Xie2020,Liu2025CPTN}), reflecting a continued emphasis on electricity–heat integration and a slower transition toward fully multi-vector, cross-sectoral formulations.

At the network-modeling level, electricity networks are generally represented via AC/DC power flow or linearized DistFlow models, often coupled with convex relaxations, such as SOCP, to retain tractability (~\cite{10378966,Xu2020,Liu2019,ALIZADEH2024122162,Qin2019}). Heat networks are commonly formulated under CF-VT assumptions, using nodal energy balances and simplified thermal loss relations; dynamic effects (e.g., transport delays, thermal inertia) are frequently approximated via quasi-dynamic models or neglected (~\cite{Wu2023a,Chen2023b,Liu2019}), although variable-flow operation (often with constant temperature) is also considered (~\cite{Hu2024}). Gas networks are typically modeled using steady-state Weymouth pressure–flow relationships, with linepack dynamics included only in quasi-dynamic or PDE-based formulations (~\cite{Zhao2021,Zhang2023,Dou2020}). Integration across energy carriers is commonly realized through energy hubs (~\cite{Zhao2021,Haghighi2015,Wang2019,Zhou2020}). Spatial scopes range from building clusters (~\cite{Lu2023,Cesena2020,Lu2021,Lv2022}) and smart districts (~\cite{9596598,Lu2021,Ji2024,Gao2022}) to regional/transmission-level systems (~\cite{ZHANG2024123593,Han2023,Wang2019,Zhang2023}), with sampling intervals typically spanning minutes to hours. For resilience at distribution level, a two‑stage framework links a steady‑state gas network and a distribution power network through gas‑fired generation and electricity‑driven pumps, coordinating pre‑event mobile storage placement with post‑event restoration (reconfiguration and crew routing)~\cite{Wang2025_ResilientIEGDS_MES}. 

\paragraph{Architectures}

\begin{figure*}[t]
\centering
\includegraphics[width=0.85\textwidth]{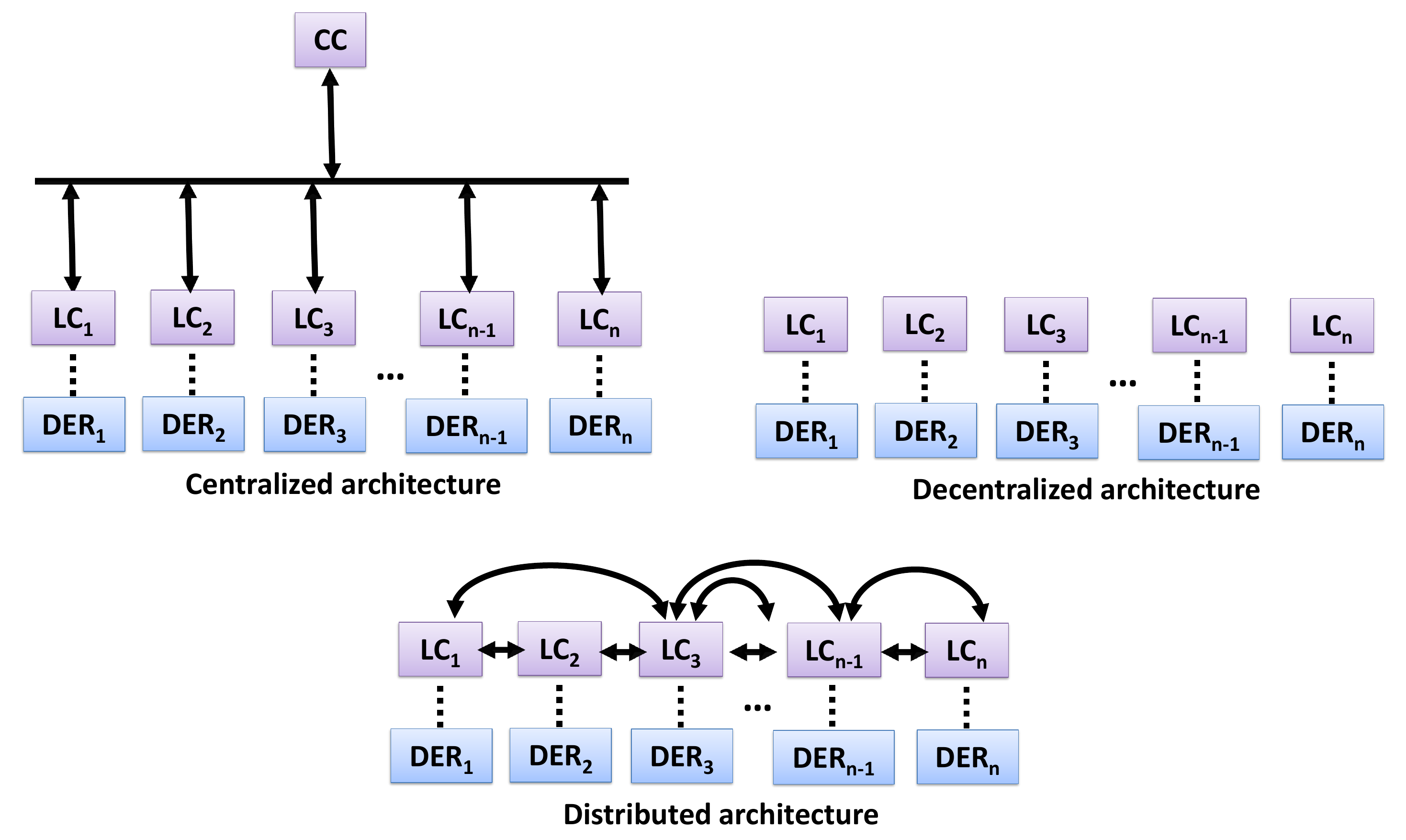}
\caption{Centralized, distributed and decentralized architectures. LC = Local Controller; DER = Distributed Energy Resource; CC = Central Coordinator. Centralized offers system‑wide visibility but can be a scalability and resilience bottleneck; distributed reduces computation/privacy burdens via coordination but incurs communication/consensus overhead; decentralized enables autonomy with no global guarantees on coupled network constraints.}
\label{Architecture}
\end{figure*}

Figure~\ref{Architecture} illustrates three decision‑making architectures, centralized, distributed, and decentralized, showing how information and control responsibilities are allocated across local controllers and any coordinating entity.

Centralized schemes provide full system visibility but lack scalability and robustness against communication failures; distributed architectures enable coordinated optimization via information exchange while reducing computational and privacy burdens; whereas decentralized schemes operate solely on local information, offering autonomy but lacking guarantees on global feasibility or stability.

The architectures across the referenced works predominantly align with centralized structures (e.g.,~\cite{Cesena2019,Zhang2024,Hu2024,Guo2022,Zhou2020,Liu2019,Chen2023,Chen2022}). In contrast, distributed and decentralized architectures use problem decomposition to improve scalability, reduce single‑point bottlenecks, limit information sharing, and enable parallel computation (~\cite{Xu2020,Chen2023b,Zhang2023,9743661}). Decentralized and distributed architectures appear in fewer works, usually leveraging Alternating Direction Method of Multipliers (ADMM)-based decomposition or consensus schemes; they often lack formal guarantees of convergence and optimality. Some studies implement fully decentralized or distributed schemes, such as ~\cite{Xu2020}, which achieves full distribution without any coordinator through iterative consensus. Reinforcement learning (RL) approaches also reflect these patterns, with~\cite{Li2023} adopting centralized training and decentralized execution, and~\cite{9596598} combining decentralized actors with central critics.

Hierarchical architectures~\cite{Jia2024,Jin2021,Deng2020}, where decision-making is organized across distinct layers, each with different roles and scopes, are commonly adopted in IES to manage complexity, multi-energy coupling and different time scales. These frameworks often follow a bi-level structure, where the upper level focuses on global objectives such as maximizing operator revenue or system-wide efficiency, while the lower level optimizes local decisions to minimize user costs, sometimes considering thermal comfort through indicators like Predictive Mean Vote (PMV) (~\cite{Li2022}). Some studies include centralized upper levels for global coordination with decentralized lower levels~\cite{Li2022,10252152,Zhang2023}, fully distributed upper levels~\cite{Liu2022}, and tri-level frameworks where intermediate coordination uses ADMM~\cite{Gan2021}. 

\begin{table*}[t]
\centering
\footnotesize
\setlength{\tabcolsep}{3.5pt}
\renewcommand{\arraystretch}{1.2}
\caption{Model, network, and architecture used in the references.}
\label{tab:model_network_architecture}

\begin{adjustbox}{width=\textwidth}
\begin{tabular}{|P{3.2cm}|P{1.25cm}|P{1.25cm}|P{1.15cm}|P{1.45cm}|P{1.05cm}|P{1.05cm}|P{1.25cm}|P{1.35cm}|P{1.9cm}|}
\hline
\textbf{References} &
\multicolumn{2}{c|}{\textbf{Model type}} &
\multicolumn{4}{c|}{\textbf{Network type}} &
\multicolumn{3}{c|}{\textbf{Architecture}} \\
\hline
& \textbf{SS} & \textbf{Dyn.}
& \textbf{Elec.} & \makecell{\textbf{Heat/}\\\textbf{Cool}} & \textbf{Gas} & \textbf{Other}
& \textbf{Cent.} & \textbf{Hier.} & \makecell{\textbf{Dist./}\\\textbf{Dec.}} \\
\hline

\cite{10378966,Li2023,Cao2018,9596598,9743661} &  & \cmark & \cmark & \cmark &  &  &  &  & \cmark \\ \hline
\cite{Xu2020,Coelho2021,Chen2023b,Wang2024,Han2024,Han2023} &  & \cmark & \cmark & \cmark & \cmark &  &  &  & \cmark \\ \hline
\cite{Jia2024,Liu2022,10252152} &  & \cmark & \cmark & \cmark &  &  &  & \cmark & \cmark \\ \hline
\cite{Li2022} &  & \cmark &  & \cmark &  &  &  & \cmark & \cmark \\ \hline
\cite{Deng2020} & \cmark &  & \cmark & \cmark &  &  & \cmark & \cmark &  \\ \hline
\cite{ZHANG2024123593} &  & \cmark & \cmark &  & \cmark &  & \cmark & \cmark &  \\ \hline
\cite{Gan2021} & \cmark &  & \cmark &  & \cmark &  &  &  & \cmark \\ \hline
\cite{Chen2023,Jin2021} &  & \cmark & \cmark & \cmark &  &  & \cmark & \cmark &  \\ \hline
\cite{Zhang2023} &  & \cmark & \cmark & \cmark & \cmark &  &  & \cmark & \cmark \\ \hline
\cite{Xie2020} & \cmark &  & \cmark &  & \cmark & \cmark & \cmark &  &  \\ \hline
\cite{Wang2018} & \cmark &  & \cmark &  & \cmark &  &  &  & \cmark \\ \hline
\cite{Zhao2021a} & \cmark &  & \cmark &  & \cmark &  & \cmark &  &  \\ \hline
\cite{Haghighi2015,Duan2024} & \cmark &  & \cmark & \cmark & \cmark &  & \cmark &  &  \\ \hline
\cite{Zhou2020} & \cmark &  & \cmark & \cmark &  &  & \cmark &  &  \\ \hline
\cite{Liu2025CPTN} &  & \cmark & \cmark &  &  & \cmark & \cmark & \cmark  &  \\ \hline
\cite{Wang2023a} &  & \cmark &  & \cmark &  &  & \cmark &  &  \\ \hline
\cite{Zhao2021,Huo2019,Lyu2023,Lv2022,Qin2019,Gan2024,Wang2024_DynamicEquivalentGasStorage,Wang2025_ResilientIEGDS_MES} &  & \cmark & \cmark &  & \cmark &  & \cmark &  &  \\ \hline
\cite{Zhang2024,ALIZADEH2024122162,Liu2019,Wang2019,Li2025,Chen2023a,Li2024,Zhai2020,Cesena2020,Gao2022,Dou2020,Cesena2019,Hu2024,Wu2023a} &  & \cmark & \cmark & \cmark & \cmark &  & \cmark &  &  \\ \hline
\cite{9437972,Xiao2023,Lu2021,Chen2022,Wu2023,Lu2023,Jiang2021,Guo2022,Luo2021,Tan2020,Ji2024} &  & \cmark & \cmark & \cmark &  &  & \cmark &  &  \\ \hline

\end{tabular}
\end{adjustbox}

\vspace{2pt}
\raggedright\footnotesize
\textbf{Legend:} SS = Steady–State; Dyn. = With dynamics; Elec. = Electricity; Heat/Cool = Heating/Cooling; 
Cent. = Centralized; Hier. = Hierarchical; Dist./Dec. = Distributed/Decentralized.
\end{table*}

\vspace{1.5em}

\subsubsection{Optimization Techniques}
This subsection reviews optimization and control approaches for IES operation.
\paragraph{Deterministic and Convex Techniques}
Table~\ref{tab:methods_uncertainty} focuses on the optimization and control methods used for IES operation. Mathematical optimization remains the dominant paradigm, encompassing mixed-integer linear and nonlinear programming (MILP, MINLP) (see~\cite{Tan2020,Guo2022,Qin2019}) as well as second-order cone programming (SOCP) formulations (see~\cite{Xu2020,Chen2023a,Gao2022}). Convex relaxations~\cite{Xu2020,Wang2018,Chen2023a,Wang2024,Gao2022}, such as SOCP for branch power flow and Weymouth gas equations, and approximation techniques, such as linear or piecewise linear approximations~\cite{Liu2019, Chen2023,Zhang2023,Qin2019} or inner box approximation~\cite{Li2025}, are often employed to address non-convexities and ensure tractability. One adopted approach to make the problem more tractable is to assume fixed mass flows in supply and return pipes of the DHC network~\cite{ALIZADEH2024122162, Liu2019, Wu2023}, which is inefficient when compared to variable mass flow operation~\cite{KUNTUAROVA2024132189}. 
In~\cite{Wang2024_DynamicEquivalentGasStorage} an inscribed ellipsoid is computed by \emph{semidefinite optimization} to capture per‑period gas–power coupling, and a time‑varying polyhedral “storage” envelope is tightened via \emph{mixed‑integer linear} feasibility checks; the problem remains a deterministic multi‑period dispatch and solves much faster than transient finite‑difference/finite‑volume gas models. 
\paragraph{Couplings and Uncertainty}
For coupled electricity–transport studies, a bi‑level, centralized framework combines a dynamic traffic‑assignment model (with queuing at fast‑charging and battery‑swapping stations, multiple charging/discharging power levels, and aggregate state‑of‑charge tracking) with a convexified distribution‑network operation model whose nodal prices are formed endogenously; the bi‑level is reformulated via optimality conditions and solved as a mixed‑integer linear program~\cite{Liu2025CPTN}. Another approach embeds a static traffic equilibrium (routing and charging) into a unified power–gas–transport operation model and applies a two‑stage bound‑tightening procedure to improve the convex approximation and solution quality~\cite{Xie2020}.

Uncertainty is considered in approximately half of the studies in Table~\ref{tab:methods_uncertainty}, often related to renewable generation and load variability (e.g.,~\cite{Han2023,Hu2024}), but also covering aspects such as reward mechanisms~\cite{9596598}, line attacks~\cite{ZHANG2024123593}, communication delays~\cite{9743661}, market prices~\cite{Liu2019}, line outages~\cite{Chen2023}, and operational conditions~\cite{Cesena2020}. 
Some studies explicitly model uncertainty through stochastic programming (~\cite{ALIZADEH2024122162,Li2025,Han2023,Huo2019,Lyu2023}), robust optimization (~\cite{Zhai2020,Lu2023,Chen2023b,Deng2020}), Info-Gap theory (~\cite{Lu2023}), or distributionally robust approaches using ambiguity sets (i.e., set of distributions that could plausibly describe the uncertainties) (~\cite{Zhao2021,Li2025,Zhou2020}) or Conditional Value at Risk (CVaR)-based formulations (~\cite{ALIZADEH2024122162,Li2025,Liu2019}). Scenario-based methods such as Latin hypercube sampling (~\cite{Zhai2020,Ji2024,Han2023,Lyu2023}) and Wasserstein metrics (~\cite{Zhao2021,Li2025,Han2023}) are common for renewable variability and load uncertainty, although scalability and conservatism are seldom analyzed. These techniques are often coupled with decomposition approaches such as ADMM (~\cite{Gan2021,Xu2020,Chen2023b,Coelho2021,Jia2024}), Lagrangian relaxation (~\cite{Gan2021,Jia2024}), and column-and-constraint generation (~\cite{Lu2023,Chen2023,Gao2022}). Scenario‑decomposition with a selective progressive‑hedging scheme is used to accelerate two‑stage restoration with mobile storage and crew routing, considering damage scenarios after extreme events and accounting for travel/repair times and load profiles within the restoration horizon~\cite{Wang2025_ResilientIEGDS_MES}.
\paragraph{Objectives and Performance Metrics}
Typical optimization objectives aim to minimize total operational costs, which may include fuel expenses, import/export costs, and penalties for CO$_2$ emissions (~\cite{Wang2024,Lu2023,9596598}).
Other objectives also involve profit-sharing based on game-theoretical methods, as described in~\cite{10252152}, or maximizing social welfare and reward (~\cite{Xiao2023,Liu2019,Liu2022}), as well as penalizing voltage deviations (~\cite{Zhao2021,Li2023}). In restoration settings, objectives often penalize unserved energy and late repairs under network and resource constraints, as in two‑stage formulations coordinating reconfiguration, crew routing, gas‑pump supply, and mobile storage~\cite{Wang2025_ResilientIEGDS_MES}.
A few studies adopt multi-objective optimization approaches to balance conflicting goals. For instance, trade-offs are sought between cost minimization and voltage deviation reduction, or between maximizing economic benefits and enhancing node flexibility (~\cite{Chen2023a,Chen2022,Li2024}). These multi-objective frameworks often adopt methods ranging from stochastic formulations solved via Normal Boundary Intersection, to interval-based models using $\epsilon$-constraint and fuzzy decision techniques for Pareto front selection. 
\paragraph{Meta-heuristics and Game-Theoretic Approaches}
Meta-heuristics serve as flexible alternatives when the dimensionality and complexity of the  problem renders exact optimization impractical. Methods like genetic algorithms (~\cite{ZHANG2024123593,Li2022}), quantum evolutionary algorithms (~\cite{Deng2020}), fruit fly optimization (~\cite{Wang2019}), and teaching-learning-based optimisation (TLBO) (~\cite{Haghighi2015}) are used to explore non-convex spaces or upper-level decision layers in bi-level settings. Despite their versatility, these approaches are heuristic by nature, offering limited guarantees on convergence or constraint satisfaction, and often depend on user-defined parameters and termination rules.

Game-theoretic frameworks are proposed to capture strategic interactions among agents and stakeholders. Cooperative games, such as Nash Bargaining (~\cite{Wang2024,ALIZADEH2024122162,Han2023}), and dynamic non-cooperative (~\cite{Duan2024}) or Stackelberg or Markov games (~\cite{Li2022,Li2023}), are used for benefit allocation, transactive control, and coordination across networks. These methods provide a sound theoretical basis for decentralized decision-making and incentive compatibility, though equilibrium computation remains challenging in high-dimensional or nonlinear settings.

\begin{table*}[t]
\centering
\footnotesize
\setlength{\tabcolsep}{3.5pt}
\renewcommand{\arraystretch}{1.2}

\caption{Methods and uncertainty sources used in the references.}
\label{tab:methods_uncertainty}

\begin{adjustbox}{width=\textwidth}
\begin{tabular}{|P{3.1cm}|P{1.55cm}|P{1.55cm}|P{1.55cm}|P{1.35cm}|P{1.25cm}|P{1.9cm}|P{1.25cm}|P{1.1cm}|}
\hline
\textbf{References} &
\multicolumn{5}{c|}{\textbf{Method}} &
\multicolumn{3}{c|}{\textbf{Uncertainty}}\\ \hline
& \textbf{MOpt} & \textbf{MH} & \textbf{RL} & \textbf{GT} & \textbf{FB} & \textbf{RES} & \textbf{Load} & \textbf{Other}\\
\hline
\cite{Xiao2023} &  & \cmark & \cmark &  & \cmark & \cmark &  &  \\ \hline
\cite{Li2024}   &  &        & \cmark &  & \cmark & \cmark & \cmark &  \\ \hline
\cite{9596598}  &  &        & \cmark &  & \cmark & \cmark & \cmark & \cmark \\ \hline
\cite{Hu2024}   &  &        & \cmark &  & \cmark & \cmark &  &  \\ \hline
\cite{Li2023}   &  &        & \cmark & \cmark & \cmark & \cmark & \cmark & \cmark \\ \hline
\cite{Duan2024} &  & \cmark &        & \cmark &        &        &        &        \\ \hline
\cite{Cao2018,Wang2024,Liu2022,Gan2024,Wang2024_DynamicEquivalentGasStorage} & \cmark & & & \cmark & & & & \\ \hline
\cite{Luo2021}  & \cmark &  &  & \cmark &  &  & \cmark &  \\ \hline
\cite{Han2023}  & \cmark &  &  & \cmark &  & \cmark & \cmark &  \\ \hline
\cite{ALIZADEH2024122162,Li2022} & \cmark & & & \cmark & & \cmark & & \\ \hline
\cite{Haghighi2015,Wang2019} &  & \cmark & & & & & & \\ \hline
\cite{Deng2020} &  & \cmark & & & & \cmark & & \\ \hline
\cite{ZHANG2024123593} & \cmark & \cmark & & & & & & \cmark \\ \hline
\cite{Guo2022}  & \cmark & \cmark & & & & & & \\ \hline
\cite{Han2024}  &  &  &  &  & \cmark &  &  &  \\ \hline
\cite{Xu2020,Wang2018,Xie2020,Lu2021,Coelho2021,Wu2023,Cesena2019,Wang2023a,Lv2022,Jia2024,Wu2023a,Liu2025CPTN} & \cmark & & & & & & & \\ \hline
\cite{10378966,9437972,10252152,Jin2021,Dou2020} & \cmark & & & & \cmark & & & \\ \hline
\cite{Zhang2024,Qin2019,Jiang2021,Chen2023b,Zhang2023} & \cmark & & & & & \cmark & \cmark & \\ \hline
\cite{Zhai2020,Zhou2020,Zhao2021a,Lyu2023,Gao2022,Gan2021,Tan2020,Ji2024,Zhao2021,Huo2019} & \cmark & & & & & \cmark & & \\ \hline
\cite{Lu2023,Chen2023a} & \cmark & & & & & \cmark & \cmark & \cmark \\ \hline
\cite{9743661,Liu2019,Chen2022,Chen2023,Cesena2020,Wang2025_ResilientIEGDS_MES} & \cmark & & & & & & & \cmark \\ \hline
\cite{Li2025}  & \cmark & & & & & & \cmark & \\ \hline
\end{tabular}
\end{adjustbox}

\vspace{2pt}
\raggedright\footnotesize
\textbf{Legend:}
MOpt = Mathematical Optimization; MH = Meta-heuristics; RL = Reinforcement Learning; GT = Game Theory; FB = Feedback; 
RES = Renewable Energy Sources; Load = Load uncertainty; Other = other uncertainty sources.
\end{table*}

\subsubsection{Feedback Control Strategies for IES}
As shown in Table~\ref{tab:methods_uncertainty}, most existing studies adopt open-loop approaches, which lack the inherent robustness of feedback, closed-loop mechanisms. Such mechanisms are essential for mitigating unknown disturbances and unmodeled dynamics, which become increasingly relevant under operational uncertainty. While open-loop frameworks may suffice for day-ahead scheduling, they are inadequate for intra-day operation and real-time control, where adaptability to evolving conditions is critical. To address these limitations, advanced feedback-based strategies such as MPC and RL have emerged as promising alternatives (see Box 2). These approaches are discussed in the following sections.

\begin{figure*}[!t]              
  \centering
    \begin{tcolorbox}[
  enhanced,
  width=\textwidth,
  title=Box 2: Model Predictive Control and Reinforcement Learning,
  colback=lightblue, colframe=blue!40!black,
  colbacktitle=lightblue_title, coltitle=black,
  fonttitle=\bfseries, boxrule=0.5pt, arc=4pt,
  left=6pt, right=6pt, top=6pt, bottom=6pt,
  breakable=false              
]
Model Predictive Control (MPC)~\cite{rawlings2017mpc} and Reinforcement Learning (RL)~\cite{GONG2023121740} are two advanced optimization-based methods that share conceptual and structural similarities, especially when applied to dynamic systems and viewed through the lens of feedback control. Classical MPC is formulated as a receding-horizon optimal control problem, where at each time step a finite-horizon optimization problem is solved using a model of the system dynamics, typically in the form \( x(t+1) = f(x(t), u(t)) \). The control inputs \( \{u(t), \dots, u(t+N-1)\} \) are chosen to minimize a cost function of the form \( J = \sum_{t=0}^{N-1} \ell(x(t), u(t)) + \ell_f(x(N)) \), subject to system dynamics and constraints on states and controls. Only the first input is applied, and the process repeats at the next time step with updated measurements, incorporating feedback implicitly. MPC is traditionally model-based using first-principles or linear models, but can also be data-driven, employing system identification or learned models such as neural networks.

In contrast, RL is generally regarded as a learning-based decision-making technique. It is concerned with learning a policy \( \pi^*(x) \) that maximizes expected cumulative reward, formalized through the Bellman optimality equation:
\[
V^*(x) = \max_u \mathbb{E}\left[r(x, u) + \gamma V^*(x')\right],
\]
where \( V^* \) is the optimal value function, \( r(x, u) \) is the reward, and \( \gamma \in (0,1] \) is a discount factor. RL can be model-free, learning directly from interaction data, or model-based, where a learned or given model of the environment is used for planning and policy improvement. 

Despite their differing formulations, MPC and RL can both be interpreted as feedback control strategies in the context of dynamic systems. MPC explicitly incorporates state measurements in its control loop through repeated re-optimization, while RL policies are typically state-dependent mappings, \( u = \pi(x) \), providing closed-loop control. 
\end{tcolorbox}
\end{figure*}

\paragraph{Model Predictive Control}

Studies such as~\cite{10378966,9437972,10252152,Jin2021,Dou2020} adopt MPC approaches in a deterministic context. In~\cite{10252152}, a hierarchical MPC framework is developed to optimally coordinate an integrated energy system. At the upper level, an MPC uses reduced-order energy models to manage power exchanges among multi-energy subsystems. Meanwhile, at the lower level, decentralized MPC regulators locally control the subsystems with faster sampling intervals. 
Recent works~\cite{FALSONE2019141,9347679} introduce decentralized and fully distributed approaches for large-scale mixed-integer linear problems using primal or dual decomposition with adaptive constraint tightening, offering finite-time feasibility and suboptimality bounds. Drawing inspiration from~\cite{FALSONE2019141}, a decentralized MPC scheme with an energy coordinator is proposed in~\cite{10378966} to optimally manage the operation of an IES district, ensuring finite-time feasibility for the MPC problem. 

Despite these advances, the stability and persistent feasibility of existing MPC frameworks for IES are not analyzed or guaranteed. The underlying problem is a large-scale, mixed-integer nonlinear optimization problem, which is NP-hard and involves coupling across constraints and decision variables. Guaranteeing stability and recursive feasibility, often through terminal constraints or cost elements, further complicates the design, particularly under uncertainty. These challenges remain inadequately addressed. 

\noindent\textit{Feasibility and Stability.}
\emph{Recursive feasibility} refers to the property that if the optimization problem is feasible at the current time step, then it remains feasible at all subsequent time steps when the first optimal control action is applied and the optimization is solved again in a receding horizon manner. This property is crucial in practice, as loss of feasibility may lead to constraint violations or the inability to compute a control input. Ensuring recursive feasibility typically requires the design of terminal constraint sets that are positively invariant and compatible with system dynamics and operational constraints. \emph{Stability} concerns the convergence of the closed-loop system to a desired equilibrium point or trajectory under the MPC law. In deterministic settings, stability is commonly established by demonstrating that the optimal cost function acts as a Lyapunov function, which decreases at each time step. This usually requires the inclusion of a suitable terminal cost and terminal region that satisfy certain control-invariance and decrease conditions. In stochastic or uncertain settings, stronger notions such as robust or probabilistic stability may be required, which significantly increase computational and theoretical complexity. See~\ref{NMPC} for further details.

\noindent\textit{Challenges and Future Directions.}
For large-scale, networked systems such as IES, stability and recursive feasibility of the MPC scheme become considerably more challenging to establish. The presence of discrete decisions, nonlinear dynamics, inter-temporal couplings, and network constraints complicates the construction of invariant terminal sets and Lyapunov functions. Moreover, uncertainty in renewable generation, load demand, and market prices further undermines classical MPC stability proofs.

Current research largely focuses on developing tractable formulations of the MPC problem for IES. A promising direction is the integration of MPC with learning-based techniques, such as machine learning or RL. These hybrid approaches aim to combine the predictive capabilities of MPC with the adaptability of learning methods, such as differentiable MPC embedded within RL algorithms or RL-inspired policy learning integrated into MPC schemes, improving robustness and scalability for complex, dynamic environments.

\paragraph{Reinforcement Learning}
Other noteworthy methods used for managing IES operations include RL and deep RL~\cite{Xiao2023,9596598,Hu2024,Li2023,Duan2024}. These are typically model-free approaches. The state space includes various IES-related states, such as storage levels, while the action space comprises setpoints and control inputs for the electricity, heating, and gas networks. The reward function is typically designed to minimize operational costs and may also incorporate rewards for maintaining network reliability. Actor-critic architectures dominate (~\cite{9596598,Li2023}), with centralized training and decentralized execution common in multi-agent scenarios. Uniquely,~\cite{Hu2024} formulates the control task as a Constrained Markov Decision Process (C-MDP) and solves it via a Safe RL method (PD-TD3), integrating soft penalties for violations in network variables such as gas pressure or power flow. In~\cite{Xiao2023} state violations are penalized, but the network is not explicitly modeled, and state/action spaces exclude network variables. 
RL methods are usually applied to sub-networks of limited size, such as microgrids, and present several challenges, including lack of interpretability, robustness, and the need for retraining whenever system changes occur. Moreover, they require careful tuning of numerous parameters to avoid under- or over-fitting, and obtaining suitable datasets for training remains a challenge. Safe RL formulations have been proposed, yet they remain computationally intensive and offer limited theoretical assurances.

\noindent\textit{RL Limitations and Rationale for Inclusion.} 
RL studies on IES operation are \emph{not network‑aware} in the sense adopted in this review: they do not model network topology, nodal balances, branch/pipe equations, or per‑carrier limits, and therefore treat feasibility only indirectly and partially (e.g., via reward shaping or simulator checks). This limitation matters at operational horizons, where spatial feasibility and time‑coupled transport (e.g., thermal advection in district heating, linepack in gas) materially constrain admissible schedules. Nevertheless, we include RL works because they act on the same operational layer and decision space, thus providing a useful algorithmic baseline, and, more importantly because
learning‑based control is a rapidly developing field; continued advances, especially toward topology, and physics‑aware formulations, may help address these network‑awareness gaps.

A practical way forward is to move from purely data‑driven, model‑free policies toward learning formulations that incorporate the network model and topology, so that decisions are made with respect to carrier physics and spatial constraints; in parallel, topology‑aware representations (e.g., graph‑based encoders) can expose nodal/edge structure and local limits to the policy. For the electrical distribution layer, the tutorial~\cite{Glover2025_DRLDistribution} captures the spatial and network aspects through a simulator-in-the-loop design: each action is evaluated by a distribution power-flow solver (OpenDSS), and the agent observes network states (e.g., bus voltages, power flows, device and switch statuses), so feasibility reflects the feeder’s topology and physics. The learning policies themselves are model-free, and the study is simulation-based rather than purely data-driven; importantly, it remains electrical-only and does not extend to multi-energy networks (district heating or gas).

\subsection{A Conceptual MPC Scheme for IES Operation Management}~\label{NMPC}

In this section, we present a conceptual MPC scheme that incorporates system dynamics, DER interactions, storage management, and network constraints

\begin{figure}[t]
\centering
\includegraphics[width=0.5\textwidth]{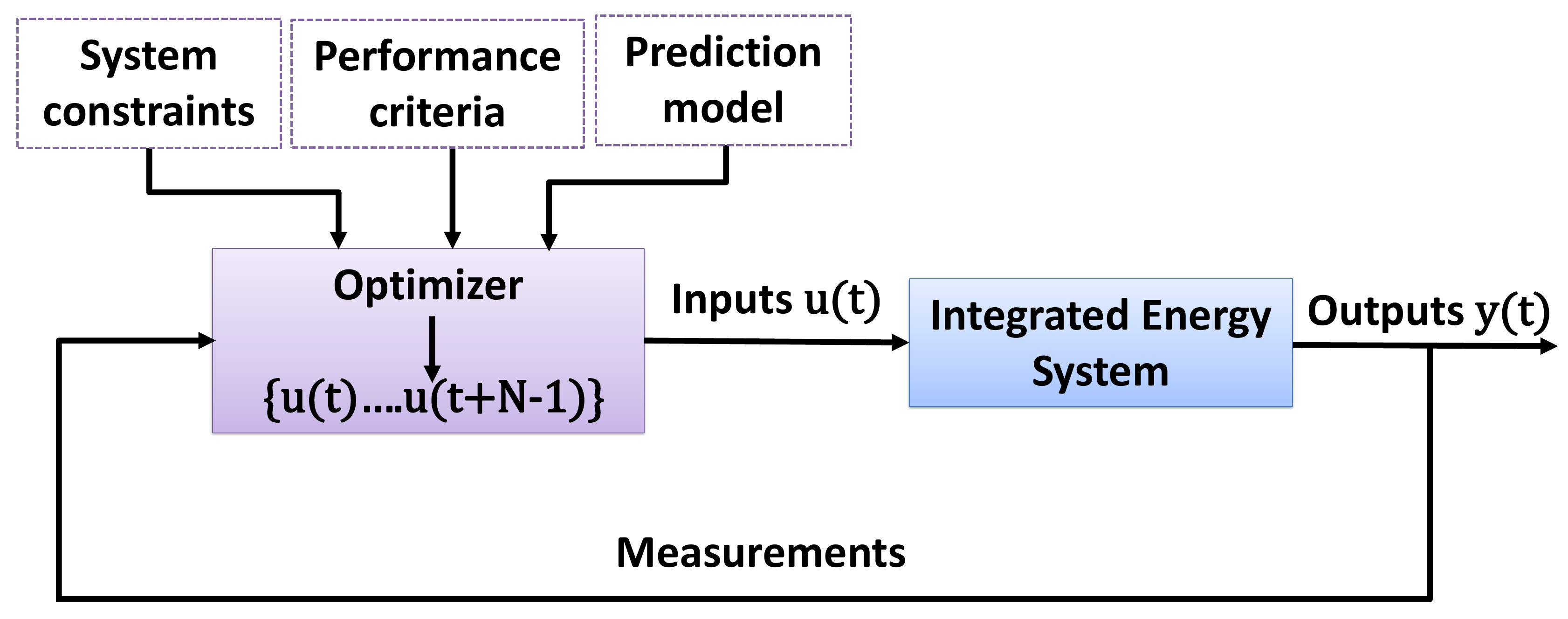}
\caption{Model Predictive Control architecture.}
\label{MPCArchitecture}
\end{figure}

We consider the network‑aware IES model of Sec.~\ref{MEN_model}, whose carrier‑network dynamics (district heating/cooling and gas) and algebraic constraints are compactly written in~\eqref{eq:dynamic},~\eqref{eq:ss}, and are coupled with device- and building‑level models connected at network nodes. Figure~\ref{MPCArchitecture} provides a schematic overview of the MPC architecture applied to this IES, illustrating the control structure: measurements and forecasts enter the prediction model and the optimizer; a constrained, finite‑horizon optimization is solved; the first optimal control action is applied to the IES. The diagram highlights the feedback loop between the physical IES and the optimizer, emphasizing the role of MPC in enforcing network‑coupled constraints while adapting to evolving conditions.

The objective of MPC is to minimize a cost function $J$ over the prediction horizon $N$:
\begin{equation*}
\min_{\{\bm{u}(t)\}_{t=0}^{N-1}} 
J = \sum_{t=0}^{N-1} \ell(\bm{x}(t), \bm{u}(t)),
\end{equation*}
where $\ell(\bm{x}(t), \bm{u}(t))$ is the stage cost function, which typically aggregates fuel or energy purchase costs, start-up and shut-down charges, efficiency losses, and deviations from desired operating set-points.

To improve closed-loop properties, typically a terminal cost $\Phi(\bm{x}(N))$ can be included, yielding
\begin{equation*}
J = \sum_{t=0}^{N-1} \ell(\bm{x}(t), \bm{u}(t)) + \Phi(\bm{x}(N)).
\end{equation*}
The terminal cost $\Phi(\bm{x}(N))$ approximates the optimal infinite-horizon cost-to-go from the terminal predicted state onward. Additionally, a terminal set $\mathcal{X}_f \subseteq \mathbb{R}^n$ can be imposed as a constraint
\begin{equation*}
\bm{x}(N) \in \mathcal{X}_f,
\end{equation*}
where $\mathcal{X}_f$ is designed to be positively invariant under a given local control law. That is, for every $\bm{x} \in \mathcal{X}_f$, there exists a control input that keeps the system state within $\mathcal{X}_f$ for all future time steps while satisfying the constraints. Proper design of $\Phi(\cdot)$ and $\mathcal{X}_f$ allows the optimal cost to serve as a Lyapunov function, thereby ensuring closed-loop stability and recursive feasibility of the MPC scheme in the nominal case~\cite{rawlings2017mpc}.

In practical applications, especially in large-scale and uncertain systems, strict (hard) enforcement of state and input constraints may lead to infeasibility of the optimization problem. To mitigate this issue, constraints are often relaxed through the introduction of slack variables $\epsilon(t) \ge 0$, leading to so-called \emph{soft constraints}. For instance, a hard constraint of the form
\begin{equation*}
g(\bm{x}(t), \bm{u}(t)) \le 0
\end{equation*}
can be reformulated as
\begin{equation*}
g(\bm{x}(t), \bm{u}(t)) \le \epsilon(t), 
\quad \epsilon(t) \ge 0,
\end{equation*}
where $\epsilon(t)$ represents the amount of constraint violation. The slack variables are penalized in the objective function, typically using an $\ell_1$- or $\ell_2$-norm penalty:
\begin{equation*}
J = \sum_{t=0}^{N-1} \ell(\bm{x}(t), \bm{u}(t)) 
+ \Phi(\bm{x}(N)) 
+ \rho_\epsilon \sum_{t=0}^{N-1} \|\epsilon(t)\|_1,
\end{equation*}
where $\rho_\epsilon > 0$ is a sufficiently large penalty parameter. 

Soft constraints guarantee that the optimization problem remains feasible even when disturbances, modeling errors, or conflicting constraints would otherwise render it infeasible. If the penalty $\rho_\epsilon$ is chosen sufficiently large, constraint violations occur only when strictly necessary. When combined with appropriate terminal ingredients, the MPC formulation ensures recursive feasibility by construction and contributes to closed-loop stability of the overall MPC scheme.

The MPC optimization problem for IES is subject to constraints of this form, with $t = 0, \ldots, N-1 $:
\begin{align*}
&\bm{x}(0) = \bm{x}_0, \\
&\text{system model}, \eqref{eq:dynamic}, \eqref{eq:ss}, \quad t = 0, \ldots, N-1, \\
&\bm{x}_{\min} \leq \bm{x}(t) \leq \bm{x}_{\max},\\
&\bm{u}_{\min} \leq \bm{u}(t) \leq \bm{u}_{\max}, \\
&\bm{y}_{\min} \leq \bm{y}(t) \leq \bm{y}_{\max}, \\
&\bm{h}(\bm{x}(t), \bm{u}(t), \bm{d}(t)) \leq 0, 
\end{align*}
where:
\begin{itemize}
    \item $\bm{x}_0$ is the initial state;
    \item $\bm{x}_{\min}, \bm{x}_{\max}$ are the state constraints;
    \item $\bm{u}_{\min}, \bm{u}_{\max}$ are the input constraints;
    \item $\bm{y}_{\min}, \bm{y}_{\max}$ are the output constraints;
    \item $\bm{h}(\cdot)$ denotes any additional nonlinear constraints.
\end{itemize}
The state vector, $\bm{x}(t)$, represents the states of the system at time $t$, e.g., energy levels in storage units or temperature in thermal systems. The control vector, $\bm{u}(t)$, represents the control inputs, e.g., power generation levels, pump/compressor powers, mass/volume flows. The output vector $\bm{y}(t)$ collects network and device quantities of interest (e.g., pipe/supply temperatures). A disturbance vector can be added, representing external disturbances, e.g., energy demands, renewable energy generation. Nonlinear constraints, $\bm{h}(\cdot)$, can represent models and specific operating constraints of relevant DERs, such as battery storage systems and heat pumps, as well as buildings, e.g., $h_{\mathrm{DER}}(x(t),u(t),\delta(t),m(t))\le 0$ and $h_{\mathrm{build}}(x(t),u(t))\le 0$, evaluated at time \(t\).

In this setting, the MPC needs to include multiple types of constraints, a mix of continuous and integer constraints. In addition to storage devices discussed in Section~\ref{DYN}, illustrative examples of mixed-integer constraints are provided below~\cite{10378966,Parisio2014MPCMicrogrid}.

\paragraph*{Example of mixed-integer structure.} \mbox{}\\
\textit{(i) Binary/mode variables.}
\newline
These binary vectors implement device logic (on/off, mode selection) used by the constraints above and below:
\[
m(t)\in\{0,1\}^{n_m},\quad
\delta(t)\in\{0,1\}^{n_\delta},
\]
where
\begin{itemize}
  \item $m(t)$: mode‑selection vector (e.g., heating/cooling, operation regimes),
  \item $\delta(t)$: on/off status across units.
\end{itemize}

\medskip
\textit{(ii) On/off with minimum up/down times.}
\newline
These constraints model the on/off status of a unit and enforce minimum up/down times by requiring the unit to remain on/off for at least $T_i^{\text{up}}/T_i^{\text{down}}$ periods after a switching event.

Minimum up/down times are enforced directly on the status variable via mixed‑integer linear inequalities. For a unit $i$ with status $\delta_i(k)\in\{0,1\}$

\[
\begin{aligned}
\delta_i(k)-\delta_i(k-1) &\le \delta_i(\tau),\\[-1mm]
&\forall\, \tau=k+1,\dots,k+T_i^{\text{up}}-1 \ \text{ with }\ \tau\le T,
\end{aligned}
\]
\[
\begin{aligned}
\delta_i(k-1)-\delta_i(k) &\le 1-\delta_i(\tau),\\[-1mm]
&\forall\, \tau=k+1,\dots,k+T_i^{\text{down}}-1 \ \text{ with }\ \tau\le T.
\end{aligned}
\]

so that, after a switch on (off), the unit must remain on (off) for at least $T^{\text{up}}_i$ ($T^{\text{down}}_i$) periods.

\medskip
\textit{(iii) Mode selection for reversible heat pumps (heating/cooling).}
\newline
These constraints enforce mutually exclusive heating/cooling modes and relate delivered heat/cooling to the electrical input via COP(coefficients of performance):
\[
Q^{\mathrm{heat}}_{\mathrm{HP}}(t)\le m_{\mathrm{heat}}(t)\,\overline{Q}^{\mathrm{heat}},\quad
Q^{\mathrm{cool}}_{\mathrm{HP}}(t)\le m_{\mathrm{cool}}(t)\,\overline{Q}^{\mathrm{cool}},
\]
\[
m_{\mathrm{heat}}(t)+m_{\mathrm{cool}}(t)\le 1,\quad
m_{\mathrm{heat}}(t),\, m_{\mathrm{cool}}(t)\in\{0,1\},
\]
\[
P_{\mathrm{el,HP}}(t)=\frac{Q^{\mathrm{heat}}_{\mathrm{HP}}(t)}{\mathrm{COP}^{\mathrm{heat}}(t)}
+\frac{Q^{\mathrm{cool}}_{\mathrm{HP}}(t)}{\mathrm{COP}^{\mathrm{cool}}(t)},
\]
where
\begin{itemize}
  \item $Q^{\mathrm{heat}}_{\mathrm{HP}}(t),\,Q^{\mathrm{cool}}_{\mathrm{HP}}(t)$: delivered heat/cooling at time $t$.
  \item $\overline{Q}^{\mathrm{heat}},\,\overline{Q}^{\mathrm{cool}}$: heat/cooling capacities (upper bounds).
  \item $m_{\mathrm{heat}}(t),\, m_{\mathrm{cool}}(t)\in\{0,1\}$: mutually exclusive mode indicators (heating or cooling).
  \item $P_{\mathrm{el,HP}}(t)$: electrical input power of the heat pump at time $t$.
  \item $\mathrm{COP}^{\mathrm{heat}}(t),\, \mathrm{COP}^{\mathrm{cool}}(t)$: (possibly time‑varying) coefficients of performance.
\end{itemize}

The MPC problem described here represents a large-scale MINLP. Because of this complexity, hierarchical and distributed control architectures are often explored to decompose the problem, reduce computational burden, and improve scalability.

\begin{remark}
  Given the mixed‑integer structure, nonlinear multi‑carrier physics, and network couplings, computing a positively invariant terminal set $\mathcal{X}_f$ and a terminal cost $\Phi(\cdot)$ that certify recursive feasibility and closed‑loop stability is particularly challenging; standard control‑invariance and Lyapunov arguments become difficult to construct and verify at realistic scales.  
\end{remark}

\subsection{Discussion}

Existing IES studies, as summarized in Tables~\ref{tab:model_network_architecture} and~\ref{tab:methods_uncertainty}, commonly rely on simplified, steady-state and convexified network models, whose relaxations can be inaccurate in the presence of nontrivial power losses, voltage drops, and hydraulic/thermal transients, thereby producing schedules that may violate physical and operational limits and risk infeasibility in practice. This challenge is further exacerbated by the fact that feasibility and optimality are not rigorously established, with very limited analysis of algorithmic guarantees across the literature. Scalability is likewise under-examined: most works confine evaluation to small, district scale cases or coarse time resolutions, and only a handful explicitly assess scalability via simulations or complexity analysis~\cite{10378966,Jiang2021,Chen2022}. 
For instance,~\cite{10378966} reports up to three orders of magnitude runtime reduction in simulations,~\cite{9743661} provides an effort estimate tied to local variable dimensionality but without formal complexity bounds,~\cite{Chen2022} observes significant rising computation times as systems grow, and~\cite{Jiang2021} finds that fine temporal resolution impedes scaling to larger networks. Without comprehensive scalability analysis, the applicability of existing approaches to large-scale, real-world IES remains uncertain.

The majority of references provide limited or no rigorous analysis of key algorithmic properties such as feasibility, convergence, and optimality, with only a few exceptions offering formal results. While~\cite{10378966} discusses finite-time feasibility and performance based on prior results,~\cite{10252152} addresses feasibility pragmatically and proves uniqueness of a bargaining solution under certain conditions,~\cite{9743661} establishes exponential rate convergence with optimality for its algorithm, and~\cite{Han2024} provides a stability analysis ensuring asymptotic convergence in distributed control. Several works (e.g.,~\cite{Jia2024,Xu2020}) discuss performance, convergence behavior, or robustness in simulation, only for specific case studies, without formal proof or clarity on implementation details. Overall, there is a significant gap in thorough theoretical analysis across the literature, which raises concerns about the reliability, scalability, and practical applicability of the proposed methods under real-world conditions.

Handling uncertainty continues to be a significant gap, with robust and stochastic formulations for uncertainty insufficiently explored. Although two-stage stochastic, robust, and distributionally robust formulations aim to improve decision-making under uncertainty, existing studies fail to critically assess the feasibility of the solutions obtained, particularly because an approximate, deterministic problem is ultimately solved. It is not examined whether robust approaches introduce excessive conservatism or whether stochastic methods yield infeasible solutions to the original stochastic problem. The risk that solutions derived from sampled scenarios may fail when exposed to out-of-sample variability is overlooked.
Computation time is a fundamental constraint for grid-support services and coordination of distributed flexibility. In MPC, solutions must be computed within the sampling interval; as model fidelity and system size increase to district scale, centralized MILP/MINLP formulations become intractable. This reality highlights the need for distributed optimization methods; however, the field remains still dominated by centralized architectures.

\section{Open Challenges and Outlook}~\label{Challenges} 

The following discussion outlines the major challenges that hinder the optimization and control of integrated energy systems and highlights emerging research directions.

\subsection{Open Challenges}

\paragraph{Physics-aware modeling}
IES couple multiple energy carriers (electricity, heat, gas, water) with distinct dynamics and time constants: electricity balances instantaneously while heat/gas exhibit inertia (thermal mass, linepack). Using steady-state surrogates biases load and dispatch. Evidence from district heating shows that neglecting thermal transients, pipeline inertia, and mixing yields biased peak load estimation and suboptimal operation, with demand underestimation up to 20\%—whereas dynamic models improve load prediction and dispatch feasibility~\cite{Massrur2018}. Likewise, steady-state gas models fail to capture time varying pressures and flow delays, producing unrealistic dispatch schedules for gas-fired units. By including dynamic behavior, both heating and gas networks enable more accurate load predictions, improved system stability, and enhanced feasibility of coordinated energy dispatch~\cite{Massrur2018a}. 

The choice of temporal discretization (e.g., 15 min vs. hourly) and spatial detail (component level vs. aggregated zones) is another critical factor. Higher resolution better captures short-term dynamics and storage effects, but increases dimensionality and computational burden. Coarser models reduce run time yet risk missing transient phenomena and misrepresenting constraints. 

Furthermore, operational dispatch for DERs is often inherently complex due to binary decisions combined with nonlinear efficiency curves. The resulting MINLP formulations are particularly challenging at fine temporal resolutions. Common strategies, such as convex relaxations, piecewise-linear approximations, and surrogate models, can improve tractability but often compromise accuracy or feasibility in tight operating regimes. Ensuring feasibility, therefore compliance with physical constraints, while reducing complexity remains an open methodological problem. 

\paragraph{Scalable optimization}
Even with improved models, operational IES optimization remains large‑scale and tightly coupled. Nonlinear network physics across carriers, binary device decisions, and intertemporal constraints create problems that are provably hard to solve; relaxations and linearizations help, but can lose accuracy or feasibility where flows, pressures, and temperatures are constrained by equipment and safety limits. Most formulations stay centralized and are tested on modest networks or coarse time grids, with limited or no analysis of convergence, optimality, and, crucially, real‑time feasibility. To address this, researchers employ decomposition techniques (e.g., Benders, ADMM, Lagrangian relaxation), model reduction (e.g., simplified hydraulics, reduced-order thermal networks), and problem-specific heuristics to partition the problem into tractable subproblems with near-optimal solutions. However, these approaches still lack robustness and formal guarantees, and their scalability for real-time operation remains largely unexplored.

\paragraph{Learning \textit{with} networks, not just \textit{from} data.}
There is growing interest in learning-based control for operational decision-making at minutes-to-hours horizons. This has value for scalability and adaptation, but methods that treat the system as a black box, without making the learning process aware of the physical network and its topology, are ill-suited to IES operation. In multi-carrier settings, valid schedules depend on the physics of each carrier (for example, power-flow relations in electricity and thermal or hydraulic transport in district heating and gas) and on spatial constraints at nodes and along connections. 
A clear direction, therefore, is to design learning schemes that explicitly incorporate the network model and the system topology, so that decisions are taken with respect to these physical and spatial limits. 

Concretely, one route is to expose nodal and line structure through graph‑based representations so the learning algorithm operates on the actual grid topology and local limits; recent reviews document how graph neural models leverage connectivity to improve prediction and control in power networks~\cite{Liao2022GNNReview}. An alternative route is to keep the relevant carrier solvers in the loop during training and evaluation, so every candidate action is evaluated through the governing equations rather than by penalty alone, as proposed in~\cite{Glover2025_DRLDistribution} for the distribution power network. Constraint‑aware learning and safe RL address the need to respect operating limits (e.g., voltages, pressures, temperatures, ratings) at decision time rather than only in expectation; a recent survey~\cite{Su2025SafeRL} summarizes mechanisms to incorporate safety criteria directly against the (electric) power network model—using power‑flow equations and device limits with simulator‑in‑the‑loop evaluation—so that actions are judged by their feasibility on the electrical topology.

Physics‑informed learning further reduces reliance on large datasets by embedding physical laws or model structure directly into training objectives, yielding enhanced sampling efficiency and better generalizability in power‑system applications~~\cite{ZHANG2023113688}. 

Taken together, these directions move the field beyond black‑box policies toward network‑aware formulations that remain directly tied to carrier physics and to spatial feasibility, which we see as a concrete and necessary research direction for integrated energy systems.

\paragraph{Interoperability and cyber‑physical challenges}
Real-world IES deployments involve heterogeneous stakeholders (power, gas, heat operators) with fragmented regulations, data silos, and misaligned incentives. Decentralized or distributed optimization methods that respect privacy and autonomy are essential, yet standards and interfaces for secure data exchange remain immature. Privacy-preserving approaches (e.g., federated optimization, limited-information protocols) show promise but require organizational and regulatory alignment to unlock system-level benefits.

\begin{figure*}[t]
\centering
\includegraphics[width=\textwidth]{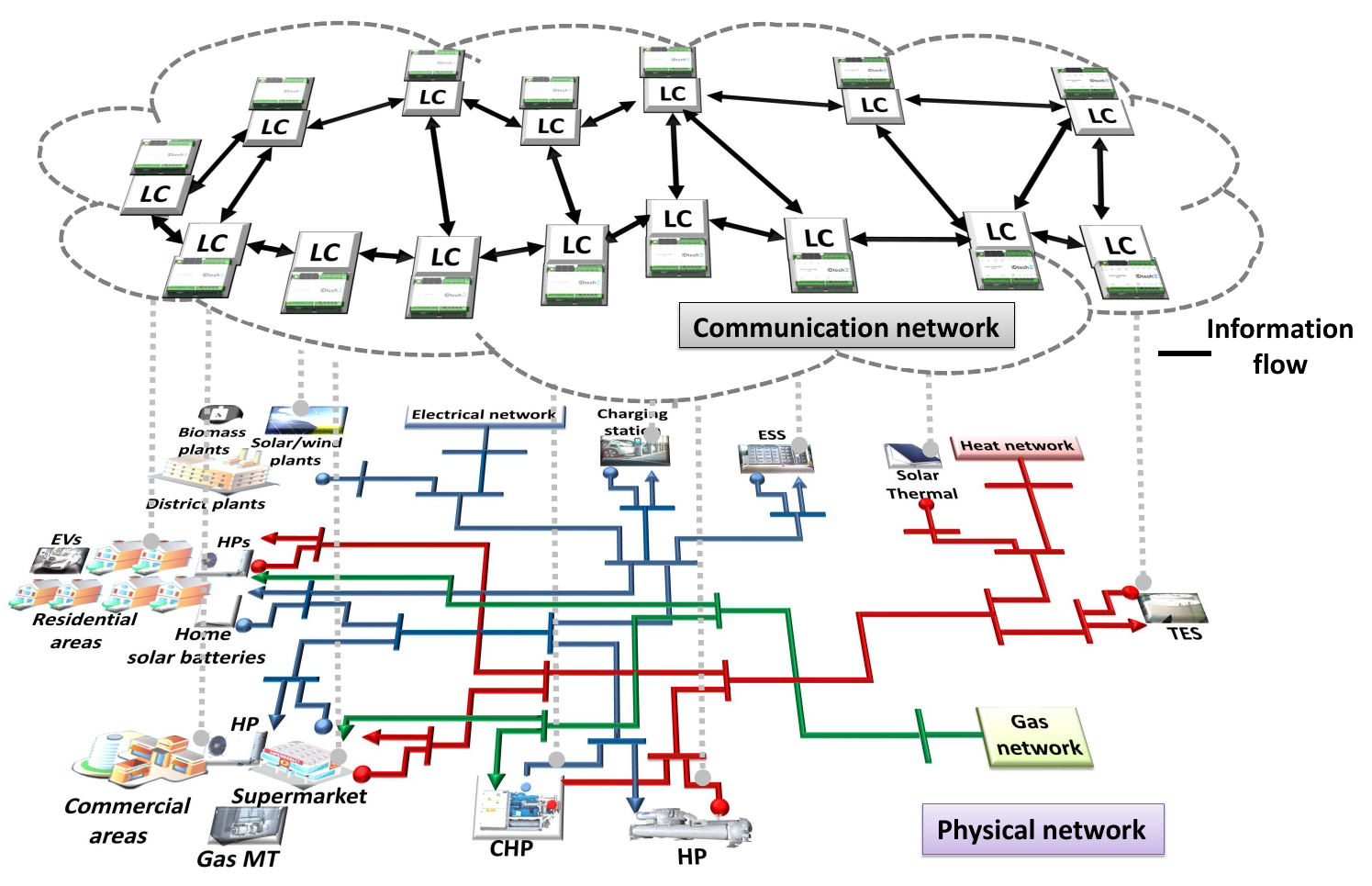}
\caption{Cyber‑physical view of an integrated energy system. A communication layer enables coordination among distributed devices but introduces latency, reliability, data quality, and security constraints that affect feasible real‑time operation—motivating latency‑aware MPC, resilient estimation/control, and robust architectures.}
\label{MEN_cloud}
\end{figure*}

A recent review of digitized grid practice~\cite{Xie2022_MassivelyDigitizedGrid} highlights that, even within the electric sector, data sharing is constrained by confidentiality, uneven sensor coverage, and heterogeneous data management stacks.
Increasing digitalization through IoT, edge control, and digital twins further transforms IES into cyber‑physical systems. Figure~\ref{MEN_cloud} illustrates this cyber‑physical perspective, highlighting the coupling between the physical infrastructure and the communication layer. Each device has a local controller (LC) that optimizes its operation and communicates with neighboring controllers via a local communication network, enabling coordinated decision-making and data exchange. Operation then depends not only on the physical energy networks but also on communication networks that introduce latency, packet loss, data quality issues, and cybersecurity risks. These factors constrain feasible control actions and must be considered when designing network‑aware optimization and real‑time control. In practice, this calls for latency‑aware MPC, attack‑resilient estimation and control, and architectures that remain robust when communication performance degrades.

While this aspect is set to become increasingly important in the future, it has not  received attention in existing research so far. 

Reproducibility and benchmarking also remain underdeveloped. High‑fidelity component and network models, particularly for thermal and gas infrastructures, are scarce, limiting rigorous cross‑comparison. There is a shortage of public, high‑fidelity component/network models for IES, especially for thermal and gas infrastructures, and a lack of representative open datasets for abnormal or extreme regimes. Even for the (electric) power grid, widely used AI studies rely on small test systems and synthetic scenarios whose sampling can be unrepresentative, leading to biased conclusions~\cite{Xie2022_MassivelyDigitizedGrid}.
Open datasets, standardized test cases, shared code repositories, and formalized performance metrics (e.g., feasibility rates, violation probabilities, and cost and emissions impacts) are needed to accelerate method maturation and ensure alignment with operational requirements.

\paragraph{Uncertainty}
Operational decisions in IES are sensitive to errors in demand, renewable generation, user behavior, and market prices. While robust and stochastic optimization frameworks exist, they can be overly conservative or computationally slow. Uncertainty in IES is not only technical but also social in nature, and IES can be viewed as \emph{social–cyber–physical systems}: human behavior, comfort preferences, occupancy patterns, and user responses to incentives all introduce variability that is difficult to capture with traditional forecasting or uncertainty sets. Social acceptance and participation levels further determine whether flexibility resources (e.g., demand response, building loads, EV fleets) actually materialize, affecting the feasibility of operational plans.
Advanced approaches such as distributionally robust optimization, chance constraints, scenario reduction, and online/receding horizon control, which update decisions as information arrives, are promising. However, practical deployment requires well-defined uncertainty sets, which reflect not only physical variability but also behavioral and socio-economic drivers, integration of fast solvers, and methods that deliver computationally efficient guarantees on feasibility and performance under uncertainty, which remains critical.

\paragraph{Market integration}
Most market arrangements remain carrier-specific, so prices do not reflect cross-carrier synergies and constraints. Co-optimized or coupled market designs are needed to price electricity–heat–gas interactions consistently in time and space, avoid conflicts between carriers, and reveal marginal costs that support coordinated flexibility. Achieving this will require new products and pricing schemes that account for temporal coupling and network limits across carriers, as well as regulatory adjustments that allow multi-energy settlements to coexist with existing electricity and gas frameworks. 

\paragraph{Demand-side flexibility}
Flexible demand from buildings, thermal inertia, demand response, and electric vehicles can lower cost and emissions but is hard to coordinate at scale, particularly in an IES context where flexibility must interact consistently with electricity, heating, and gas networks. Models must respect comfort constraints, represent user behavior and device logic, and remain tractable within system‑level, multi‑carrier optimization. Progress depends on building and device models that are simple enough for real‑time use yet accurate enough to protect comfort and equipment, and on control architectures that can aggregate many small resources while enforcing network limits across carriers, ensuring that local flexibility contributes reliably to IES operation.

\subsection{Emerging Directions and Research Agenda}
Current IES optimization frameworks remain computationally demanding, with limited scalability and no formal guarantees for real-time operation. Meaningful progress requires models and algorithms that improve fidelity without sacrificing tractability, and that can guarantee feasible operation under realistic temporal and spatial resolutions. Physics-aware convexification techniques can enhance model fidelity and retain the essential network behavior while preserving tractability, and keeping the problem solvable within the time frames required for operational use. 

Scalability also depends on optimization methods that can coordinate decisions across infrastructure layers, devices, and time intervals without relying on a single central solver. Distributed and hierarchical schemes are promising in this respect, but they must be equipped with guarantees so that feasibility, convergence, and computational performance can be preserved when the system size grows. Sub-optimality can offer a pathway to scalable coordination, as well as warm-starting, and problem decompositions aligned with the physical structure of IES. Leveraging operational and historical data can further support these approaches by enabling accurate parameter estimation, refining uncertainty characterization, and informing learning-based components for adaptive decision-making. Machine learning can improve forecasts, or provide informed initializations for optimization. Reinforcement learning may eventually offer adaptive policies for complex environments, but its deployment depends on ensuring that feasibility with respect to network limits is always maintained. MPC provides an effective framework for enforcing operational constraints. Hybrid MPC–RL schemes are particularly promising for managing multi-timescale dynamics and uncertainty, combining predictive control with adaptive learning to improve real-time performance. 
In addition, consistent evaluation on larger test cases, with transparent reporting of runtime and feasibility under increased resolution, is needed to understand how these methods perform beyond small demonstrators.

Overall, a clear research trajectory lies in integrating dynamic system modeling, distributed decision-making, and learning-based adaptability within unified optimization–control architectures that span planning, operational, and real-time layers. Because data-driven and adaptive control strategies remain at an early stage, future IES frameworks must evolve to be not only scalable and robust but also theoretically grounded and capable of reliably coordinating distributed flexibility at operational timescales.

\paragraph{Key messages for practitioners.} At minutes–hours horizons, network‑aware modeling of electricity, heating, and gas is essential for feasible schedules; convex relaxations are powerful but must be used where assumptions hold; distributed/hierarchical decision architectures are needed for scale yet should come with basic guarantees (feasibility/convergence); MPC offers systematic enforcement of network and operational constraints, while learning‑based tools can enhance adaptability and forecasting; together they are effective when learning remains consistent with the network physics.

\section{Conclusions} \label{Conclusions}
This review analyzes network‑aware modeling, optimization, and control methods for Integrated Energy Systems at minutes‑to‑hours operational horizons. We highlighted where explicit network physics and topology in electricity, heating, and gas are necessary for feasible operation; where common relaxations and linearizations are effective; and where they risk loss of fidelity or guarantees. We also surveyed decision architectures and solution approaches—ranging from centralized MILP/MINLP and SOCP relaxations to distributed/hierarchical coordination and feedback control—clarifying their scalability and the scarcity of formal feasibility, stability, and convergence results in realistic settings. Looking ahead, meaningful progress requires physics‑aware yet tractable formulations, distributed optimization with guarantees, and learning methods used in support of, rather than in place of, model‑based control. Advancing uncertainty treatment, interoperability, and cyber‑physical robustness will be key to deploying practical, low‑carbon, large‑scale IES.

\bibliographystyle{IEEEtran}

\bibliography{IESref}

\end{document}